\documentclass[journal,twoside,web]{ieeecolor}

\usepackage{generic}
\usepackage[noadjust]{cite}
\usepackage{multirow}
\usepackage{numprint}
\usepackage[font=footnotesize]{caption}

\usepackage[english]{babel}
\usepackage{pifont}   % Dingbats
\usepackage{textcomp} % nice greek alphabet

\usepackage{amsmath}
\usepackage{amssymb}
\usepackage{amsfonts}
\usepackage{mathtools}
\usepackage{siunitx}
\usepackage{numprint}
\usepackage{cases}

\usepackage{subcaption}
\usepackage{graphicx}
\usepackage{xcolor}
\usepackage{rotating}
\usepackage[framemethod=TikZ]{mdframed}
\definecolor{blueNavy}{RGB}{0,64,122}

\usepackage{booktabs}
\usepackage{tabularx}
\usepackage{longtable}
\usepackage{multirow}
\usepackage{tcolorbox}
\usepackage{tabularx}
\tcbuselibrary{skins,breakable}

\usepackage{tikz}
\usetikzlibrary{external,automata,trees,shadows,matrix,fit,decorations,positioning,shapes,plotmarks,patterns,arrows,calc,quotes,tikzmark,arrows.meta}

\newcounter{theo}[section] 
\renewcommand{\thetheo}{\arabic{section}.\arabic{theo}}

\newenvironment{definitionblock}[2]{%
		\tcolorbox[colback=white,colframe=blueNavy!95,fonttitle={\bfseries},title=#1,phantomlabel=#2,left=2pt,top=1pt, bottom=1pt,right=2pt]}%
{\endtcolorbox}

\usepackage{algorithm,algorithmic}
\floatname{algorithm}{\bfseries Algorithm}
\newcommand{\KwIn}[1]{\textbf{Input:} #1}
\newcommand{\KwOut}[1]{\textbf{Output:} #1}

\makeatletter
\newcommand\fs@booktabsruled{%
	\def\@fs@cfont{\bfseries\strut}\let\@fs@capt\floatc@ruled
	\def\@fs@pre{\hrule height\heavyrulewidth depth0pt \kern\belowrulesep}%
	\def\@fs@mid{\kern\aboverulesep\hrule height\lightrulewidth\kern\belowrulesep}%
	\def\@fs@post{\kern\aboverulesep\hrule height\heavyrulewidth\relax}%
	\let\@fs@iftopcapt\iftrue
}
\makeatother

\floatstyle{booktabsruled}
\restylefloat{algorithm}
\newcommand{\mat}[1]{\ensuremath{{\mathbf{\MakeUppercase{#1}}}}}

\makeatletter
\renewcommand{\vec}[1]{%
	\ifcat\relax\noexpand#1%
	\ensuremath{\boldsymbol{\lowercase{#1}}}%
	\else
	\ensuremath{\mathbf{\lowercase{#1}}}%
	\fi
}
\makeatother
\newcommand{\sumlim}[2]{\ensuremath{\sum\limits_{#1}^{#2}}}
\makeatother

\makeatletter

\renewcommand{\iff}{%
	\Leftrightarrow
}

\newcommand{\transpose}[1]{\ensuremath{{#1}^{\textsc{t}}}}

\newcommand{\inverse}[1]{\ensuremath{{#1}^{-1}}}

\newcommand{\trace}[1]{\ensuremath{\text{Tr}\left(#1\right)}}

\newcommand{\R}{\ensuremath{\mathbb{R}}}

\newcommand{\norm}[1]{\left|\left|#1\right|\right|}

\usepackage{pifont}

\usepackage{optidef}

\newcommand\scalemath[2]{\scalebox{#1}{\mbox{\ensuremath{\displaystyle #2}}}}
\newtcolorbox{mymathbox}[1][]{colback=white, sharp corners, #1}

\def\Minus{\texttt{-}}

\def\BibTeX{{\rm B\kern-.05em{\sc i\kern-.025em b}\kern-.08em
		T\kern-.1667em\lower.7ex\hbox{E}\kern-.125emX}}
\makeatletter
\let\NAT@parse\undefined
\makeatother
\usepackage{hyperref}

\title{
	Stimulus-Informed Generalized Canonical Correlation Analysis for Group Analysis of Neural Responses to Natural Stimuli
}

\author{Simon Geirnaert, Yuanyuan Yao, Tom Francart, and Alexander Bertrand, \IEEEmembership{Senior Member, IEEE}
	\thanks{This work is supported by a junior postdoctoral fellowship fundamental research from the Research Foundation Flanders (FWO) (for S. Geirnaert - 1242524N), FWO project nr. G081722N, the European Research Council (ERC) under the European Union’s Horizon 2020 Research and Innovation Programme (grant agreement No 802895), and the Flemish Government (AI Research Program).}% <-this % stops a space
	\thanks{S. Geirnaert, Y. Yao, and A. Bertrand are with KU Leuven, Department of Electrical Engineering (ESAT), STADIUS Center for Dynamical Systems, Signal Processing and Data Analytics and with Leuven.AI - KU Leuven institute for AI, Kasteelpark Arenberg 10, B\Minus3001 Leuven, Belgium (e-mail: simon.geirnaert@esat.kuleuven.be, yuanyuan.yao@esat.kuleuven.be, alexander.bertrand@esat.kuleuven.be).}%
	\thanks{T. Francart and S. Geirnaert are with KU Leuven, Department of Neurosciences, Research Group ExpORL, Herestraat 49 box 721, B\Minus3000 Leuven, Belgium and with Leuven.AI - KU Leuven institute for AI (e-mail: tom.francart@kuleuven.be).}%
}

\begin{document}
	
	\maketitle
	
	\begin{abstract}
		Various new brain-computer interface technologies or neuroscience applications require decoding stimulus-following neural responses to natural stimuli such as speech and video from, e.g., electroencephalography (EEG) signals. In this context, generalized canonical correlation analysis (GCCA) is often used as a group analysis technique, which allows the extraction of correlated signal components from the neural activity of multiple subjects attending to the same stimulus. GCCA can be used to improve the signal-to-noise ratio of the stimulus-following neural responses relative to all other irrelevant (non-)neural activity, or to quantify the correlated neural activity across multiple subjects in a group-wise coherence metric. However, the traditional GCCA technique is stimulus-unaware: no information about the stimulus is used to estimate the correlated components from the neural data of several subjects. Therefore, the GCCA technique might fail to extract relevant correlated signal components in practical situations where the amount of information is limited, for example, because of a limited amount of training data or group size. This motivates a new stimulus-informed GCCA (SI-GCCA) framework that allows taking the stimulus into account to extract the correlated components. We show that SI-GCCA outperforms GCCA in various practical settings, for both auditory and visual stimuli. Moreover, we showcase how SI-GCCA can be used to steer the estimation of the components towards the stimulus. As such, SI-GCCA substantially improves upon GCCA for various purposes, ranging from preprocessing to quantifying attention.
	\end{abstract}
	
	\section{Introduction}
	\label{sec:intro}
	\noindent
	Traditionally, brain-computer interface (BCI) and other neuroscience applications are oriented towards active paradigms, requiring the active participation of a user following instructions. Furthermore, they use multi-trial designs, requiring the same stimulus to be repeated multiple times to be able to average the brain responses to enhance the signal-to-noise (SNR) ratio. Moreover, synthetic stimuli such as flickering checkerboard patterns or beep sounds are used to elicit more controllable and deterministic neural responses, such as the P300-response or steady-state visual-evoked potentials (SSVEPs)~\cite{nicolasAlonso2012brain}. While such controlled BCI paradigms are valuable from a scientific point of view, their practical impact is often limited to a few niche applications, for example, to re-establish communication for patients suffering from locked-in syndrome~\cite{nicolasAlonso2012brain,lotte2018review}. To open up BCI technology to much more widespread usage in the daily life context, the limiting artificial conditions of such controlled BCI paradigms need to be alleviated.
	
	In the past few years, we have seen a surge of BCI applications that are \emph{passive}, i.e., tapping into the natural behavior of a user, \emph{single-trial}, i.e., not requiring repetitions of the stimulus, and operate on \emph{natural} sensory stimuli, such as speech/music and natural video~\cite{geirnaert2021eegBased,geirnaert2022timeAdaptive,poulsen2017eeg,dmochowski2012correlated,dmochowski2014audience,ki2016attention,dmochowski2018extracting,vanthornhout2019effect,belo2021eeg,nentwich2023semantic,yao2023identifying}. These new BCI paradigms can then be employed in much more mainstream application domains, such as hearing aids and consumer earphones~\cite{osullivan2014attentional,geirnaert2021eegBased,geirnaert2021unsupervised,geirnaert2022timeAdaptive,vanthornhout2018speech}, educational sciences~\cite{poulsen2017eeg,belo2021eeg,davidesco2021neuroscience}, neuromarketing~\cite{dmochowski2012correlated}, or virtual reality environments~\cite{delvigne2021attention}. Many of these applications involve decoding stimulus-following neural responses, for example, to quantify levels of absolute~\cite{ki2016attention,vanthornhout2019effect,roebben2023are} or selective attention~\cite{osullivan2014attentional,geirnaert2021eegBased} to a particular auditory or visual stimulus. The temporal dynamics of the stimulus then result in a so-called stimulus-following neural response (i.e., the neural tracking phenomenon~\cite{aiken2008human}), which can be decoded from different neurorecording modalities such as electroencephalography (EEG)~\cite{aiken2008human}, magnetoencephalography (MEG)~\cite{ding2012neural}, or electrocorticography (ECoG)~\cite{mesgarani2012selective}. The EEG modality is particularly interesting because it is non-invasive, low cost and highly mobile~\cite{nicolasAlonso2012brain}.
	
	However, decoding neural responses to natural stimuli is much more challenging from a signal processing perspective, given that they are much more unpredictable (resulting from the nature of the stimulus) and suffer from a very low SNR, as the targeted stimulus-following neural responses are buried under all kinds of non-neural and neural noise. Given the single-trial paradigm, averaging the responses across multiple trials is not an option anymore to deal with this extremely low SNR. Therefore, much more advanced, data-driven signal processing algorithms are required and being developed to enhance the targeted stimulus-following activity and suppress all other noise~\cite{wong2018comparison,geirnaert2021eegBased,miran2020dynamic}. In this paper, we focus on the \emph{group analysis} of such stimulus-following responses, i.e., we assume a set of synchronized neural responses to the same natural stimulus is available, e.g., from a group of subjects all attending to the same stimulus. This synchronization hypothesis roots in the previously mentioned neural tracking phenomenon of natural stimuli, where the coherent neural responses across subjects arise from the synchronized stimulus-following neural responses. Importantly, this synchronization across subjects refers only to these stimulus-locked responses that are typically earlier responses, appearing in the $\SIrange{0}{400}{\milli\second}$ post-stimulus range~\cite{aiken2008human,mesgarani2012selective,ding2012neural,osullivan2014attentional}. More complex, longer-latency responses that are, for example, related to emotional arousal or cognition are unlikely to be synchronous across subjects. This hypothesis of synchronicity for group decoding has been confirmed in various papers in literature using data-driven decoding algorithms (e.g., \cite{dmochowski2012correlated,dmochowski2014audience,ki2016attention,poulsen2017eeg,deCheveigne2019multiway,diLiberto2021accurate,yao2023identifying}). The group analysis of the stimulus-following neural responses can be a goal in itself, e.g., to decode a notion of group attention to the stimulus, as employed in educational neuroscience~\cite{poulsen2017eeg} or neuromarketing~\cite{dmochowski2012correlated}. Alternatively, the group information can be leveraged to assist the decoding of stimulus-following neural responses on an individual level, using it as, e.g., a preprocessing technique to a priori improve the SNR~\cite{deCheveigne2019multiway,diLiberto2021accurate}. 
	
	In this paper, we specifically focus on a signal processing technique called \emph{generalized canonical correlation analysis} (GCCA), the multi-view extension of the frequently used canonical correlation analysis (CCA)~\cite{carroll1968generalization,kettenring1971canonical}. CCA extracts the correlated components between two views of the same activity~\cite{hotelling1936relations}. In more traditional BCI's that, for example, decode SSVEPs, CCA is one of the most commonly used algorithms for classification by maximizing the correlation between the EEG responses and template reference signals that model the different flickering stimuli at their specific frequency (including harmonics)~\cite{chen2021adaptive}. In the context of natural stimulus-following neural responses, CCA is, for example, used to decode the speech envelope of an attended speech source from EEG~\cite{dmochowski2018extracting,decheveigne2018decoding}. In CCA, a decoder on the multi-channel EEG is then simultaneously trained with an encoder on the speech envelope to find these correlated components. The resulting correlation can then be used to decode selective attention, e.g., between two competing speech sources~\cite{geirnaert2021eegBased}. Its multi-view extension is GCCA, where the objective is to decode correlated components between more than two views of the same activity~\cite{carroll1968generalization,kettenring1971canonical}. In SSVEP-based BCIs, GCCA has been used to extract more natural reference signals by extracting the correlated components from several EEG trials containing responses to the same stimulus frequency. These learned reference signals can then be used in the previous two-view CCA method to classify new SSVEPs~\cite{zhang2014frequency}. GCCA is moreover often used when multiple EEG signals from several users are available that simultaneously attend to the same natural stimulus. GCCA can then be used to quantify attention, enhance the SNR, reduce the dimensionality, or summarize the set of EEG signals~\cite{deCheveigne2019multiway,poulsen2017eeg,dmochowski2012correlated,yao2023identifying}. An excellent tutorial paper on GCCA (there dubbed MCCA) for decoding brain responses is written by de Cheveign{\'e} et al.~\cite{deCheveigne2019multiway}.
	
	A property of GCCA in this context is that it is \emph{stimulus-unaware}: to extract the correlated components from the synchronized EEG signals, it does not assume or use any stimulus information. This is attractive in situations where the stimulus is unknown or unavailable, or when it is unknown what features of the stimulus elicit decodable neural responses. However, this stimulus-unawareness can be a disadvantage at the same time, e.g., when the stimulus is available or known, which occurs when a particular stimulus is deliberately used (e.g., in neuromarketing) or can be recorded (e.g., in hearing aids, in the classroom). Exploiting the stimulus in those situations as side information to help (e.g., in a regularization context) or steer the estimation of the correlated components across the synchronized EEG activity can then be highly beneficial, especially if we consider the very low SNR of the stimulus-following neural responses. This very low SNR is even harder to cope with when the amount of estimation data is limited, as is the case in a time-adaptive, online processing context~\cite{geirnaert2022timeAdaptive}, or when the group size is limited by the application. Therefore, the objective of this paper is to develop and analyze a \emph{stimulus-informed} GCCA (SI-GCCA) algorithm that allows taking the stimulus into account when performing a group analysis of stimulus-following neural responses. While the use of CCA to extract correlated components between (individual) neural responses and the (natural) stimulus has been successfully established (see before and, e.g., \cite{dmochowski2018extracting,decheveigne2018decoding,chen2021adaptive}), such a group analysis of multiple stimulus-following neural responses where the neural decoders are specifically steered to yield responses that are (more) coherent with the natural auditory/visual stimulus has not yet been developed. 
	
	The paper is structured as follows. In Section~\ref{sec:maxvar-gcca/corrca}, we explain the well-known MAXVAR algorithm for GCCA and its correlated component analysis (corrCA) variant. In Section~\ref{sec:si-gcca/corrca}, we derive our proposed SI-GCCA algorithm. In Section~\ref{sec:experiments}, we then describe all necessary details about the datasets and experiments to analyze the developed algorithms. The results are shown and discussed in Section~\ref{sec:results-discussion}, and conclusions are drawn in Section~\ref{sec:conclusion}.
	
	\emph{Disclaimer:} A conference precursor of this manuscript has been published in~\cite{geirnaert2023stimulusInformed}. The current manuscript contains a more extensive explanation of the developed algorithm, includes an additional variant (e.g., the corrCA and SI-corrCA algorithm), does not only focus on speech as in~\cite{geirnaert2023stimulusInformed} but also includes video, includes additional experiments, and provides additional use cases of the algorithm and a more in-depth analysis.
	
	\section{MAXVAR-GCCA/corrCA}
	\label{sec:maxvar-gcca/corrca}
	\noindent
	In canonical correlation analysis (CCA), the objective is to find the components that exhibit the highest correlation across two different views or subjects. There exist multiple options to generalize CCA to more than two views or subjects as envisaged in generalized canonical correlation analysis (GCCA), such as MAXVAR, SUMCORR, SSQCORR, \dots~\cite{carroll1968generalization,kettenring1971canonical} However, the two most popular variants are SUMCORR and MAXVAR, which both have the traditional CCA for two views as a special case~\cite{sorensen2021generalized}. SUMCORR naturally extends CCA to more than two views by maximizing the sum of pairwise correlations between the different filtered views. However, this optimization problem turns out to be NP-hard with no closed-form solution~\cite{fu2016efficient}. Therefore, a relaxation of the SUMCORR-problem is often used, the MAXVAR-problem, in which the average pairwise distance between the filtered views is minimized. Hence, MAXVAR starts from a different interpretation of the CCA problem and conveniently boils down to a closed-form solution in the form of a generalized eigenvalue decomposition (GEVD), similar to CCA~\cite{horst1961generalized,carroll1968generalization,hovine2021distributed}. An interesting property of such a GEVD is that it is invariant under a scaling of one the views in the (G)CCA problem (see Lemma I in Hassani et al.~\cite{hassani2016gevd}). Because of this attractive property of the MAXVAR-GCCA formulation (i.e., boiling down to a GEVD) and because it allows for an easier introduction of the stimulus information (see Section~\ref{sec:si-gcca/corrca}), we choose this formulation in this work. The MAXVAR-GCCA formulation for stimulus-following neural responses is introduced in Section~\ref{sec:maxvar-gcca}. In the context of a group analysis of stimulus-following neural responses~\cite{deCheveigne2019multiway}, sometimes an additional constraint is added to enforce identical neural decoders across all subjects~\cite{dmochowski2012correlated,poulsen2017eeg,dmochowski2018extracting}. GCCA with this additional constraint is often referred to as `correlated component analysis' (corrCA) and will be briefly reviewed in Section~\ref{sec:maxvar-corrCA}. The additional constraint in corrCA acts as a regularizer to limit the degrees of freedom in the model, avoiding overfitting in cases where the amount of data is limited.
	
	\subsection{MAXVAR-GCCA}
	\label{sec:maxvar-gcca}
\begin{figure*}
		\centering
		\includegraphics[width=0.85\textwidth]{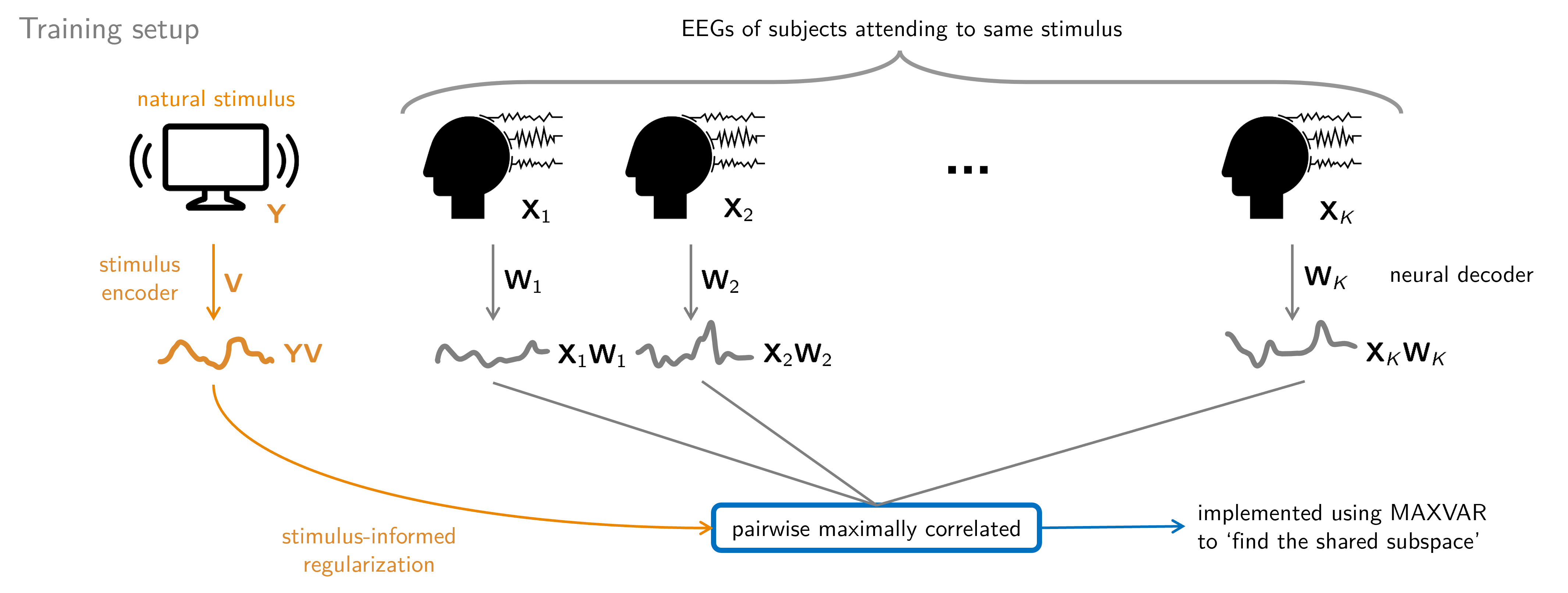}
		\caption{In this work, we consider the EEG data $\mat{X}_k$ of $K$ subjects attending to the same natural stimulus, represented by $\mat{Y}$. In GCCA, the intuitive objective is to find per-subject neural decoders $\mat{W}_k$ that maximize the pairwise correlation between the neurally decoded signals $\mat{X}_k\mat{W}_k$. Here we implement this objective via the MAXVAR framework. In our proposed SI-GCCA framework, the stimulus is included in the optimization problem via a stimulus encoder $\mat{V}$ to steer and regularize the estimation problem.}
		\label{fig:concept-sigcca}
	\end{figure*}

	\noindent
	We consider the EEG data from $K$ subjects attending to the same natural stimulus, for example, a speech signal or video. This setup is visualized in Figure~\ref{fig:concept-sigcca}. We denote $\mat{X}_k \in \R^{T \times M}$ as the EEG data of the $k$\textsuperscript{th} subject, where $T$ denotes the number of available EEG samples. Each EEG sample is $M$-dimensional, corresponding to, for example, different EEG channels and/or time-lagged copies of each channel, where the latter allows to also exploit spectral or temporal information in the data-driven decoder design. Such temporal filtering is, for example, needed to compensate for temporal differences in neural processing between subjects in the group. Given $C$ EEG channels and $L$ time lags, ranging from $-\frac{L-1}{2}$ to $\frac{L-1}{2}$ (assuming $L$ is odd), the EEG data matrix $\mat{X}_k$ when using spatiotemporal filtering will then consist of $M = CL$ columns. The resulting EEG regression matrix $\mat{X}_k$ is in that case a block Hankel matrix, e.g., for $L = 5$:
	\[
	\mat{X}_k = \begin{bmatrix}
		\mat{X}_{k,1} & \dots & \mat{X}_{k,C}
	\end{bmatrix},\]
	\[
	\mat{X}_{k,c} = \scalemath{0.81}{\begin{bmatrix}
			0 & 0 & x_{k,c}(0) & x_{k,c}(1) & x_{k,c}(2) \\
			0 & x_{k,c}(0) & x_{k,c}(1) & x_{k,c}(2) & x_{k,c}(3) \\
			x_{k,c}(0) & x_{k,c}(1) & x_{k,c}(2) & x_{k,c}(3) & x_{k,c}(4)\\
			\vdots & \vdots & \vdots & \vdots & \vdots \\
			x_{k,c}(T-3) & x_{k,c}(T-2) & x_{k,c}(T-1) & 0 & 0\\
		\end{bmatrix},}
	\]
	with $x_{k,c}(t)$ the $c$\textsuperscript{th}-channel EEG signal of the $k$\textsuperscript{th} subject. While not strictly necessary, for simplicity, we assume an equal dimensionality per subject, i.e., $M$ is the same across the EEG recordings of all subjects. We also assume that the EEG data in $\mat{X}_k$ is centered or high-pass filtered such that it is zero-mean.
	
	The objective is to find the $M \times Q$-dimensional neural decoders $\mat{W}_k \in \R^{M \times Q}$ such that the individual projected signals $\mat{X}_k\mat{W}_k \in \R^{T \times Q}$ across all subjects $k=1,\dots,K$ are on average as close as possible to each other. This can be realized by introducing a $Q$-dimensional shared signal subspace spanned by the columns of $\mat{S} = \begin{bmatrix}\vec{s}_1 & \cdots & \vec{s}_Q \end{bmatrix} \in \R^{T \times Q}$, and also optimizing for this shared signal subspace to be on average closest to $\mat{X}_k\mat{W}_k$, for all $k$~\cite{kettenring1971canonical,via2007learning,hovine2021distributed}:
	\begin{definitionblock}{MAXVAR-GCCA problem}{}
		\begin{mini}|s|
			{\substack{\mat{W}_1, \dots, \mat{W}_K,\mat{S}}}{\sumlim{k=1}{K}\norm{\mat{S}-\mat{X}_k\mat{W}_k}_F^2+\mu\sumlim{k=1}{K}\norm{\mat{W}_k}_F^2}{\label{eq:maxvar-gcca-opt}}{}
			\addConstraint{\transpose{\mat{S}}\mat{S} = \mat{I}_Q,}
		\end{mini}
	\end{definitionblock}
	\noindent with $\mat{I}_Q$ the $Q$-dimensional identity matrix and $\norm{\cdot}_F$ the Frobenius norm. The second term in~\eqref{eq:maxvar-gcca-opt} represents $\ell_2$-norm regularization or diagonal loading to avoid overfitting, where the hyperparameter $\mu$ controls the amount of regularization added. If $\mu = 0$, no regularization is used, corresponding to the original MAXVAR-GCCA problem. The constraint $\transpose{\mat{S}}\mat{S} = \mat{I}_Q$ ensures that subsequent neural decoders have orthogonal/uncorrelated outputs \emph{on average}\footnote{This is a major difference with the SUMCORR formulation, where the subsequent neural decoders for each \emph{individual} subjects should yield uncorrelated outputs.} and avoids the trivial solution where all $\mat{W}_k$'s and $\mat{S}$ are set to zero.
	
	Defining $\mat{R}_{kl} = \transpose{\mat{X}}_k\mat{X}_l \in \R^{M \times M}$ as the sample crosscorrelation matrix of $\mat{X}_k$ and $\mat{X}_l$ (autocorrelation matrix when $k = l$), 
	\[\mat{R}_{D_{xx}} = \text{Blkdiag}\!\left(\mat{R}_{11},\dots,\mat{R}_{KK}\right) \in \R^{KM \times KM}
	\]
	represents the block diagonal matrix containing the per-subject autocorrelation matrices, and 
	\[\mat{R}_{xx} = \transpose{\mat{X}}\mat{X} \in \R^{KM \times KM},\] 
	with $\mat{X} = \begin{bmatrix} \mat{X}_1 & \cdots & \mat{X}_K \end{bmatrix} \in \R^{T \times KM}$, is the correlation matrix of all EEG data, containing all correlation matrices $\mat{R}_{kl}$ in its blocks. The Karush-Kuhn-Tucker (KKT) conditions then lead to the following generalized eigenvalue problem\footnote{The derivation of \eqref{eq:gevd} is a special case of the derivation of the SI-GCCA solution in Section~\ref{sec:si-gcca}, i.e., for $\gamma = 0$.}~\cite{via2007learning,hovine2021distributed}:
	\begin{definitionblock}{MAXVAR-GCCA solution}{}
		\begin{equation}
			\label{eq:gevd}
			\left(\mat{R}_{D_{xx}}+\mu\mat{I}_{KM}\right)\mat{W} = \mat{R}_{xx}\mat{w}\mat{\Omega},
		\end{equation}
	\end{definitionblock}
	\noindent where $\mat{W} = \transpose{\begin{bmatrix}  \transpose{\mat{w}}_1 & \cdots & \transpose{\mat{w}}_K\end{bmatrix}} \in \R^{KM \times Q}$ concatenates all per-subject neural decoders. We find the optimal neural decoders $\mat{W}$ as the $Q$ generalized eigenvectors (GEVcs) corresponding to the $Q$ smallest generalized eigenvalues (GEVls), which can be found in the diagonal matrix $\mat{\Omega} \in \R^{Q \times Q}$ (see \cite{hovine2021distributed} and the derivation of SI-GCCA in Section~\ref{sec:si-gcca}). Furthermore, it can be found that
	\begin{equation}
		\label{eq:s}
		\mat{S} = \sumlim{k=1}{K}\mat{X}_k\mat{W}_k\mat{\Omega}.
	\end{equation}
	\noindent The correct scaling of the GEVcs and thus neural decoders is determined via the equality constraint $\transpose{\mat{S}}\mat{S} = \mat{I}_Q$ and \eqref{eq:s}.
	
	Note that alternatively to solving the generalized eigenvalue decomposition (GEVD) in \eqref{eq:gevd}, the neural decoders can also be found via the eigenvalue decomposition of the correlation matrix of all the pre-whitened data matrices of each subject. As such, GCCA can be formulated as concatenating two principal component analysis (PCA) blocks in a two-step analysis, which might be more intuitive and easy to understand, as explained and motivated in~\cite{deCheveigne2019multiway}. As furthermore explained in~\cite{deCheveigne2019multiway}, the inverse eigenvalues of \eqref{eq:gevd} represent the degree of correlation that exists for each component. A component that is only present in the data of a single subject corresponds to a unit eigenvalue, whereas a component shared by exactly $P$ data matrices corresponds to an associated eigenvalue of $\frac{1}{P}$. As such, the eigenvalues are smaller for signal components that are shared by many subjects.
	
	\subsection{MAXVAR-corrCA}
	\label{sec:maxvar-corrCA}
	\noindent
	In the GCCA problem in Section~\ref{sec:maxvar-gcca}, for each subject $k$, a different neural decoder $\mat{W}_k$ is trained. As an additional constraint, one could restrict these per-subject neural decoders to be the same across subjects, i.e., $\mat{W}_1 = \dots = \mat{W}_K = \mat{W} \in \R^{M\times Q}$, which is dubbed corrCA in~\cite{dmochowski2012correlated,poulsen2017eeg,dmochowski2018extracting}. From a parameter estimation point of view, this can be viewed as an additional regularization, as the number of parameters that need to be estimated drastically decreases. From a neural point of view, this assumes more uniform neural signals across subjects.
	
	In MAXVAR-corrCA with diagonal loading, the optimization problem in \eqref{eq:maxvar-gcca-opt}, therefore, boils down to:
	\begin{definitionblock}{MAXVAR-corrCA problem}{}
		\begin{mini}|s|
			{\substack{\mat{W},\mat{S}}}{\sumlim{k=1}{K}\norm{\mat{S}-\mat{X}_k\mat{W}}_F^2+\mu\norm{\mat{W}}_F^2}{\label{eq:maxvar-corrCA-opt}}{}
			\addConstraint{\transpose{\mat{S}}\mat{S} = \mat{I}_Q.}
		\end{mini}
	\end{definitionblock}
	The KKT conditions then lead to the following generalized eigenvalue problem\footnote{Again, the derivation can be viewed as a special case of the derivation in Section~\ref{sec:si-gcca}, when $\gamma = 0$.}:
	\begin{definitionblock}{MAXVAR-corrCA solution}{}
		\begin{equation}
			\label{eq:maxvar-corrca-gevd}
			\left(\sumlim{k=1}{K}\mat{R}_{kk}+\mu\mat{I}_{M}\right)\mat{W} = \left(\sumlim{k=1}{K}\sumlim{l=1}{K}\mat{R}_{kl}\right)\mat{W}\mat{\Omega}.
		\end{equation}
	\end{definitionblock}
	\noindent As opposed to the GEVD in \eqref{eq:gevd}, which is of dimension $KM$, this GEVD is only of dimension $M$, effectively showing the regularization aspect of corrCA. Moreover, the matrices in \eqref{eq:maxvar-corrca-gevd} correspond to the sum of the $M \times M$ block matrices of $\mat{R}_{D_{xx}}$ and $\mat{R}_{xx}$, respectively, showing the effect of restricting the per-subject neural decoders to be the same.
	
	\section{SI-GCCA/SI-corrCA}
	\label{sec:si-gcca/corrca}
	\noindent
	We now propose the stimulus-informed GCCA (SI-GCCA) technique for the group analysis of stimulus-following neural responses, in which we include the stimulus as side information, as visualized in Figure~\ref{fig:concept-sigcca}. The implicit assumption in SI-GCCA is, therefore, that the correlated components across the $K$ different EEG recordings correspond to (early) stimulus-following/-related neural responses, as explained in the introduction. This assumption also delineates the context in which we propose SI-GCCA, i.e., all $K$ EEG recordings are recorded using the same natural stimulus and are synchronized in time.
	
	The stimulus is incorporated in the GCCA technique with two objectives in mind. Firstly, to steer the estimation of the correlated components towards the stimulus and specific stimulus representations. Secondly, it can be viewed as task-informed regularization (in the broad sense), as it allows to take additional information into account to cope with situations where less information (e.g., little data, few subjects) is available. Both objectives will be evaluated in Section~\ref{sec:results-discussion}.
	
	\subsection{SI-GCCA}
	\label{sec:si-gcca}
	\noindent
	Apart from the $K$ zero-mean EEG signals $\mat{X}_k$, now also assume we have access to the stimulus via the $P$-dimensional stimulus representation $\mat{Y} \in \R^{T \times P}$ (see Section~\ref{sec:stimulus-features} for examples of such a representation for a speech and video stimulus). In SI-GCCA, our goal is to ensure that the shared signal subspace $\mat{S}$, which connects the EEGs of the different subjects, is also close to the stimulus representation $\mat{Y}$, i.e., the signals in $\mat{S}$ and $\mat{Y}$ should be correlated. We do this by introducing an extra term into the MAXVAR-GCCA estimation problem in \eqref{eq:maxvar-gcca-opt}, where we use a forward model/encoder $\mat{V} \in \R^{P \times Q}$ on the stimulus to map it to the shared signal subspace $\mat{S}$:
	\begin{definitionblock}{SI-GCCA problem}{}
		\begin{mini}|s|
			{\substack{\mat{W}_1, \dots, \mat{W}_K,\\\mat{V},\mat{S}}}{\sumlim{k=1}{K}\norm{\mat{S}-\mat{X}_k\mat{W}_k}_F^2+\gamma\norm{\mat{s}-\mat{Y}\mat{v}}_F^2}{\label{eq:sigcca-opt}}{}
			\breakObjective{\quad+\mu\left(\sumlim{k=1}{K}\norm{\mat{W}_k}_F^2+\norm{\mat{V}}_F^2\right)}
			\addConstraint{\transpose{\mat{S}}\mat{S} = \mat{I}_Q.}
		\end{mini}
	\end{definitionblock}
	\noindent The hyperparameter $\gamma$ determines how much weight is put onto the stimulus, i.e., how hard the stimulus `pulls' on the shared signal subspace. The motivation behind using a simple forward model $\mat{V}$ on the stimulus representation is that it essentially retains the MAXVAR-GCCA structure as in \eqref{eq:maxvar-gcca-opt}, where the stimulus acts as an additional view of the underlying shared subspace. We will show that, as a result, SI-GCCA retains the attractive property of MAXVAR-GCCA that it can be solved via a GEVD. 
	
	The Lagrangian function is equal to:
	\[
	\begin{split}
		\mathcal{L}&\!\left(\mat{W}_1,\dots,\mat{W}_K,\mat{V},\mat{S},\mat{\Lambda}\right) =(K+\gamma)\trace{\transpose{\mat{S}}\mat{S}}-\\
		&\quad 2\sumlim{k=1}{K}\trace{\transpose{\mat{S}}\mat{X}_k\mat{W}_k}+  \sumlim{k=1}{K}\trace{\transpose{\mat{W}}_k\transpose{\mat{X}}_k\mat{X}_k\mat{W}_k}-\\
		&\quad2\gamma\trace{\transpose{\mat{S}}\mat{Y}\mat{V}}+\gamma\trace{\transpose{\mat{V}}\transpose{\mat{Y}}\mat{Y}\mat{V}}+\mu\sumlim{k=1}{K}\trace{\transpose{\mat{W}}_k\mat{W}_k}+\\
		&\quad\mu\trace{\transpose{\mat{V}}\mat{V}} -\trace{\left(\transpose{\mat{S}}\mat{S}-\mat{I}_Q\right)\mat{\Lambda}}, \\
	\end{split}
	\]
	with $\mat{\Lambda} \in \R^{Q \times Q}$ a symmetric matrix containing the Lagrange multipliers. The KKT conditions then lead to the following four equations:
	\begin{numcases}{}
		\nabla_{\mat{W}_k}\left(\mathcal{L}\right) = 0 \iff \transpose{\mat{X}}_k\mat{S} = \left(\mat{R}_{kk}+\mu\mat{I}_M\right)\mat{W}_k, \; \forall k \label{eq:kkt-si-wk}\\
		\nabla_{\mat{V}}\left(\mathcal{L}\right) = 0 \iff \gamma\transpose{\mat{Y}}\mat{S} = \left(\gamma\mat{R}_{yy}+\mu\mat{I}_P\right)\mat{V}, \label{eq:kkt-si-v}\\
		\nabla_{\mat{S}}\left(\mathcal{L}\right) = 0 \iff \mat{s} = \left(\sumlim{k=1}{K}\mat{X}_k\mat{W}_k+\gamma\mat{Y}\mat{V}\right)\mat{\Omega},\label{eq:kkt-si-s}\\
		\nabla_{\mat{\Lambda}}\left(\mathcal{L}\right) = 0 \iff \transpose{\mat{s}}\mat{s} = \mat{I}_Q\label{eq:kkt-si-lambda},
	\end{numcases}    
	with $\mat{R}_{yy} = \transpose{\mat{Y}}\mat{Y} \in \R^{P \times P}$ the stimulus autocorrelation matrix and $\mat{\Omega} = \inverse{\left((K+\gamma)\mat{I}_Q-\mat{\Lambda}\right)} \in \R^{Q \times Q}$ a symmetric matrix. Define the crosscorrelation matrix between EEG data matrix $\mat{X}_k$ and stimulus data matrix $\mat{Y}$ as $\mat{R}_{ky} = \transpose{\mat{X}}_k\mat{Y} \in \R^{M \times P}$, and the augmented data matrix $\bar{\mat{X}} = \begin{bmatrix} \mat{X}_1 & \cdots & \mat{X}_K & \mat{Y} \end{bmatrix} \in \R^{T \times (KM+P)}$ and variables $\bar{\mat{w}} = \transpose{\begin{bmatrix}  \transpose{\mat{w}}_1 & \cdots & \transpose{\mat{w}}_K & \transpose{\mat{v}} \end{bmatrix}} \in \R^{(KM+P) \times Q}$, both including the stimulus data matrix and, respectively, the forward encoder. By plugging \eqref{eq:kkt-si-s} into \eqref{eq:kkt-si-wk} and \eqref{eq:kkt-si-v}, and combining all equations, we find:
	\begin{definitionblock}{SI-GCCA solution}{}
		\begin{equation}
			\label{eq:almost-gevd-sigcca}
			\left(\mat{P}\mat{R}_{D_{\bar{x}\bar{x}}}+\mu\mat{I}_{KM+P}\right)\bar{\mat{W}} = \mat{P}\mat{R}_{\bar{x}\bar{x}}\mat{P}\bar{\mat{w}}\mat{\Omega},
		\end{equation}
	\end{definitionblock}
	\noindent where $\mat{R}_{\bar{x}\bar{x}} = \transpose{\bar{\mat{X}}}\bar{\mat{X}} \in \R^{(KM+P)\times(KM+P)}$, $\mat{R}_{D_{\bar{x}\bar{x}}} = \text{Blkdiag}\!\left(\mat{R}_{11},\dots,\mat{R}_{KK},\mat{R}_{yy}\right)$, and 
	\[
	\mat{P} = \begin{bmatrix}
		\mat{I}_{KM} & \mat{0} \\
		\mat{0} & \gamma\mat{I}_P
	\end{bmatrix} \text{(weighting matrix)}.\] Using this in \eqref{eq:almost-gevd-sigcca}, we find 
	\[\mat{P}\mat{R}_{D_{\bar{x}\bar{x}}} = \text{Blkdiag}\!\left(\mat{R}_{11},\dots,\mat{R}_{KK},\gamma\mat{R}_{yy}\right),\] and 
	\[
	\mat{P}\mat{R}_{\bar{x}\bar{x}}\mat{P} = \begin{bmatrix} 
		\mat{R}_{11} & \dots & \mat{R}_{1K} & \gamma\mat{R}_{1y} \\
		\vdots & & \vdots & \vdots \\
		\mat{R}_{K1} & \dots & \mat{R}_{KK} & \gamma\mat{R}_{Ky} \\
		\gamma\mat{R}_{y1} & \dots & \gamma\mat{R}_{yK} & \gamma^2\mat{R}_{yy} \\
	\end{bmatrix}.
	\]
	
	While \eqref{eq:almost-gevd-sigcca} resembles a generalized eigenvalue problem, $\mat{\Omega}$ is not a diagonal matrix. However, it can be easily found that the underlying solution boils down to a GEVD, given that $\mat{\Omega}$ is orthogonally diagonalizable as it is a symmetric matrix:
	\begin{equation}
		\label{eq:omega-orth}
		\mat{\Omega} = \mat{U}\mat{\Sigma}\transpose{\mat{U}},
	\end{equation}
	with $\mat{U} \in \R^{Q \times Q}$ an orthogonal matrix and $\mat{\Sigma} \in \R^{Q\times Q}$ a diagonal matrix. Substituting \eqref{eq:omega-orth} in \eqref{eq:almost-gevd-sigcca} reveals the underlying GEVD that leads to the solution:
	\begin{equation}
		\label{eq:gevd-sigcca}
		\left(\mat{P}\mat{R}_{D_{\bar{x}\bar{x}}}+\mu\mat{I}_{KM+P}\right)\bar{\mat{W}}\mat{U} = \mat{P}\mat{R}_{\bar{x}\bar{x}}\mat{P}\bar{\mat{w}}\mat{U}\mat{\Sigma}.
	\end{equation}
	From \eqref{eq:gevd-sigcca}, it can be seen that the set of optimal neural decoders and stimulus encoders defined by $\bar{\mat{W}}$ are in the subspace spanned by the GEVcs from the matrix pencil $\left(\mat{P}\mat{R}_{D_{\bar{x}\bar{x}}}+\mu\mat{I}_{KM+P},\mat{P}\mat{R}_{\bar{x}\bar{x}}\mat{P}\right)$. Given that $\mat{P}$ is a symmetric matrix, and $\transpose{\left(\mat{P}\mat{R}_{D_{\bar{x}\bar{x}}}\right)} = \transpose{\mat{R}}_{D_{\bar{x}\bar{x}}}\transpose{\mat{P}} = \mat{R}_{D_{\bar{x}\bar{x}}}\mat{P} = \mat{P}\mat{R}_{D_{\bar{x}\bar{x}}}$, \eqref{eq:gevd-sigcca} is the GEVD of a matrix pencil of two real \emph{symmetric} matrices with $\mat{P}\mat{R}_{\bar{x}\bar{x}}\mat{P}$ positive definite, therefore, resulting in a real solution for GEVcs and GEVls (found on the diagonal of $\mat{\Sigma}$). Furthermore, as the solution of \eqref{eq:sigcca-opt} is only defined upon any orthogonal transformation, and since the objective function at the optimal solution can be found\footnote{Proofs are omitted for conciseness.} to be equal to $(K+\gamma)Q-\trace{\inverse{\mat{\Sigma}}}$, we find the optimal neural decoders $\mat{W}_k$ and stimulus encoder $\mat{V}$ as the $Q$ GEVcs corresponding to the $Q$ \emph{smallest} GEVls. The equality constraint in \eqref{eq:kkt-si-lambda} determines the correct scaling of the GEVcs. We summarize the SI-GCCA algorithm in Algorithm~\ref{algo:si-gcca}. A MATLAB implementation is available~\cite{geirnaert2024sigccaToolbox}.
	
	\begin{algorithm}
		\caption{SI-GCCA}
		\label{algo:si-gcca}
		\KwIn{$K$ stimulus-driven EEG signals $\mat{X}_k \in \R^{T \times M}$, stimulus features $\mat{Y}\in \R^{T \times P}$, stimulus hyperparameter $\gamma$, $\ell_2$-norm regularization hyperparameter $\mu$, subspace dimension $Q$}\\
		\KwOut{\makebox[0.1\linewidth][l]{Per-subject neural decoders $\mat{W}_k\!\in \R^{M \times Q}$}}
		%	\KwOut{Per-subject filters $\mat{W}_k\!\in \R^{M \times Q}$}
		\begin{algorithmic}[1]
			\STATE Compute correlation matrices $\mat{R}_{\bar{x}\bar{x}}$ and $\mat{R}_{D_{\bar{x}\bar{x}}} = \text{Blkdiag}\!\left(\mat{R}_{11},\dots,\mat{R}_{KK},\mat{R}_{yy}\right),$	with $\bar{\mat{X}} = \begin{bmatrix} \mat{X}_1 & \cdots & \mat{X}_K & \mat{Y} \end{bmatrix}$, $\mat{R}_{kk} = \transpose{\mat{X}}_k\mat{X}_k$, and $\mat{R}_{yy} = \transpose{\mat{Y}}\mat{Y}$
			\STATE Compute $\bar{\mat{W}}$ as the $Q$ GEVcs corresponding to the $Q$ smallest GEVls of the matrix pencil $\left(\mat{P}\mat{R}_{D_{\bar{x}\bar{x}}}+\mu\mat{I}_{KM+P},\mat{P}\mat{R}_{\bar{x}\bar{x}}\mat{P}\right)$, with $
			\mat{P} = \begin{bmatrix}
				\mat{I}_{KM} & \mat{0} \\
				\mat{0} & \gamma\mat{I}_P
			\end{bmatrix}$
			\STATE Scale the GEVcs such that $\transpose{\mat{S}}\mat{S} = \mat{I}_Q$, with $\mat{S}$ defined in \eqref{eq:kkt-si-s}
			\STATE Extract $\mat{W}_k$ from $\tilde{\mat{W}} = \transpose{\begin{bmatrix}  \transpose{\mat{w}}_1 & \cdots & \transpose{\mat{w}}_K & \transpose{\mat{v}} \end{bmatrix}}$
		\end{algorithmic}
	\end{algorithm}
	
	\subsection{SI-corrCA}
	\label{sec:si-corrCA}
	\noindent
	In the stimulus-informed corrCA (SI-corrCA) version we constrain the per-subject neural decoders to be the same across subjects. In this scenario, we are most heavily introducing additional constraints into the group decoding problem: all neural decoders are the same across subjects (corrCA) \emph{and} the correlated components must resemble the stimulus features (stimulus-informed). The optimization problem then becomes:
	\begin{definitionblock}{SI-corrCA problem}{}
		\begin{mini}|s|
			{\substack{\mat{W},\mat{V},\mat{S}}}{\sumlim{k=1}{K}\norm{\mat{S}-\mat{X}_k\mat{W}}_F^2+\gamma\norm{\mat{s}-\mat{Y}\mat{v}}_F^2}{\label{eq:sicorrca-opt}}{}
			\breakObjective{\quad+\mu\left(\norm{\mat{W}}_F^2+\norm{\mat{V}}_F^2\right)}
			\addConstraint{\transpose{\mat{S}}\mat{S} = \mat{I}_Q.}
		\end{mini}
	\end{definitionblock}
	The KKT conditions then, similarly to Section~\ref{sec:si-gcca}, lead to the following generalized eigenvalue problem:
	\begin{definitionblock}{SI-corrCA solution}{}
		\begin{equation}
			\label{eq:gevd-sicorrca}
			\begin{split}
				\left(\begin{bmatrix}
					\sumlim{k=1}{K}\mat{R}_{kk} & \mat{0}\\
					\mat{0} & \gamma\mat{R}_{yy}
				\end{bmatrix}+\mu\mat{I}_{M+P}\right)\begin{bmatrix}
					\mat{W} \\ \mat{V}
				\end{bmatrix} = \\\begin{bmatrix}
					\sumlim{k=1}{K}\sumlim{l=1}{K}\mat{R}_{kl} & \gamma\sumlim{k=1}{K}\mat{R}_{ky}\\
					\gamma\sumlim{k=1}{K}\mat{R}_{yk} & \gamma^2\mat{R}_{yy}
				\end{bmatrix}\begin{bmatrix}
					\mat{W} \\ \mat{V}
				\end{bmatrix}\mat{\Omega},
			\end{split}
		\end{equation}
	\end{definitionblock}
	\noindent
	merging properties from both the MAXVAR-corrCA solution in \eqref{eq:maxvar-corrca-gevd} and the SI-GCCA solution in \eqref{eq:gevd-sigcca}.
	
	\section{Experiments}
	\label{sec:experiments}
	\noindent
	We will evaluate and compare the different GCCA and SI-GCCA variants based on the EEG signals of a group of subjects listening to the same speech signals or watching the same videos. In this section, we describe the experiments in terms of the datasets (Section~\ref{sec:dataset}), the EEG preprocessing and stimulus feature extraction (Section~\ref{sec:preprocessing}), the performance evaluation schemes and metrics (Section~\ref{sec:performance-evaluation}), and the decoder setup (Section~\ref{sec:decoder-setup}). MATLAB code to reproduce all experiments is available online~\cite{geirnaert2024sigccaToolbox}.
	
	\subsection{Datasets}
	\label{sec:dataset}
	\noindent
	
	Two datasets are used to compare the various methods: one with natural speech and one with video footage as the stimulus.
	
	\subsubsection{Speech dataset}
	\label{sec:speech-dataset}
	The speech dataset is taken from the first experiment of Broderick et al.~\cite{broderick2018electrophysiological} and contains the EEG data of 19 normal-hearing subjects listening to the same audiobooks. The EEG data is recorded with a 128-channel BioSemi ActiveTwo system and is re-referenced to the average mastoid channel. The data was recorded per subject separately in 20 trials of around $\SI{3}{\minute}$ long and is cut into 52 $\SI{1}{\minute}$-trials for further processing. In total, there is $\SI{52}{\minute}$ of synchronized EEG/speech data per subject. This dataset is publicly available~\cite{broderick_michael_p_2019_5080270}.
	
	\subsubsection{Video dataset}
	\label{sec:video-dataset}
	The video dataset is taken from the Single-Shot dataset of Yao et al.~\cite{yao2023identifying} and contains the EEG data of 20 healthy subjects with normal or corrected-to-normal vision watching the same video footage. The video footage consists of a single moving person during a performance (e.g., dance, magic shows). The EEG data is recorded with a 64-channel BioSemi ActiveTwo system. The data was recorded per subject separately in 2 trials of around $36$ and $\SI{35}{\minute}$ and is cut into 56 $\SI{1}{\minute}$-trials for further processing. This dataset is publicly available~\cite{yao2024data}.
	
	\subsection{Stimulus feature extraction and EEG preprocessing}
	\label{sec:preprocessing}
	\noindent
	
	\subsubsection{Stimulus features}
	\label{sec:stimulus-features}
	
	\paragraph{Speech}
	The speech signals are represented by the low-frequency envelope of the speech signal, computed using the Hilbert transform~\cite{diLiberto2015low}. In various works (e.g.,~\cite{osullivan2014attentional,diLiberto2015low,vanthornhout2018speech,geirnaert2021unsupervised}), it has been shown that the EEG signals track this speech envelope. Moreover, we bandpass-filter the speech envelope using a $4$\textsuperscript{th}-order Butterworth filter in the $\delta$-band ($\SIrange{1}{4}{\hertz}$), where it is shown to give good tracking results~\cite{vanthornhout2019effect}. The resulting signal is stored in a one-dimensional vector $\vec{y} \in \R^{T}$ containing the samples of the envelope at different time instances during the experiment. This representation $\vec{y}$ of the speech stimulus will be used in the SI-GCCA framework to create the matrix $\mat{Y}$ (see also Section~\ref{sec:decoder-setup}). 
	
	Note that while in this work, we choose the speech envelope as an exemplary stimulus representation for natural speech, other (higher-level) features such as, e.g., phoneme and word onsets, phoneme and word surprisal, or cohort entropy could be (even additionally) used, as they are shown to also synchronize with neural signals~\cite{brodbeck2018rapid,gillis2021neural,zhang2023leading}.
	
	\paragraph{Video}
	The video stimulus is represented by an object-based version of the average optical flow, i.e., the magnitude of the pixel-wise velocity vector between frames averaged across all pixels belonging to an object in the video (after object segmentation)~\cite{yao2023identifying}. This again results in a one-dimensional vector $\vec{y}\in \R^{T}$. In Yao et al.~\cite{yao2023identifying}, it is shown that the object-based optical flow leads to significant tracking in the EEG signals of subjects watching a video. This object-based optical flow is computed after resampling (including anti-aliasing) the video data to $\SI{30}{\hertz}$ and resizing it to $854 \times 480$ pixels~\cite{yao2023identifying}.
	
	\subsubsection{EEG preprocessing}
	\label{sec:eeg-preprocessing}
	\noindent
	The EEG data are preprocessed similarly to the original references of the datasets~\cite{broderick2018electrophysiological,yao2023identifying}. This means that the EEG data for the video dataset is first preprocessed by interpolating bad channels, average re-referencing, notch filtering to remove the powerline noise, and regressing out eye activity using EOG. For the speech dataset, the EEG data were re-referenced to the average of the mastoid channels. Additionally, the EEG data is bandpass-filtered (between $\SIrange{1}{4}{\hertz}$ for the speech dataset and $\SIrange{0.5}{15}{\hertz}$ for the video dataset), downsampled (to $\SI{8}{\hertz}$ for the speech dataset and $\SI{30}{\hertz}$ for the video dataset), and normalized. In both cases, the EEG data are afterwards normalized per $\SI{1}{\minute}$-trial by setting the mean per channel to zero and the Frobenius norm across all channels to one.
	
	\subsection{Performance evaluation}
	\label{sec:performance-evaluation}
	
	\subsubsection{Testing procedure}
	\label{sec:testing-procedure}
	\noindent
	To investigate the influence of different variables such as group size, amount of training data, and number of channels, we perform Monte-Carlo experiments in which we fix 2 variables to a default value, and perform a sweep on the third one. 50 Monte Carlo runs are used for each value in the sweep. Possible interactions between these 3 variables will be investigated ad hoc (see Section~\ref{sec:interaction}). The default values of the different variables are $\SI{40}{\minute}$ of training data, 64 channels, and using all subjects. In the speech case, this means that the 64 channels corresponding to the 64-channel BioSemi system of the video dataset are chosen from the 128-channel EEG system of the speech dataset. As such, the baseline values between the speech and video datasets are the same. When $\SI{40}{\minute}$ of training are selected, the rest of the trials are split into $25\%$ validation set (for hyperparameter estimation, see Section~\ref{sec:hyperparameter-selection}) and $75\%$ test set. Per Monte Carlo run, the training trials are randomly sampled from all available trials. When varying the group size or number of channels, the chosen group or channels are similarly randomly changed between Monte Carlo runs. The same training-validation-test set split per Monte Carlo run is used for all methods, such that the results are directly comparable between methods.
	
	The test window length, over which the correlation performance metrics below are computed, is $\SI{60}{\second}$.
	
	\subsubsection{Evaluation metrics}
	\label{sec:evaluation-metrics}
\begin{figure*}
		\centering
		\includegraphics[width=0.85\textwidth]{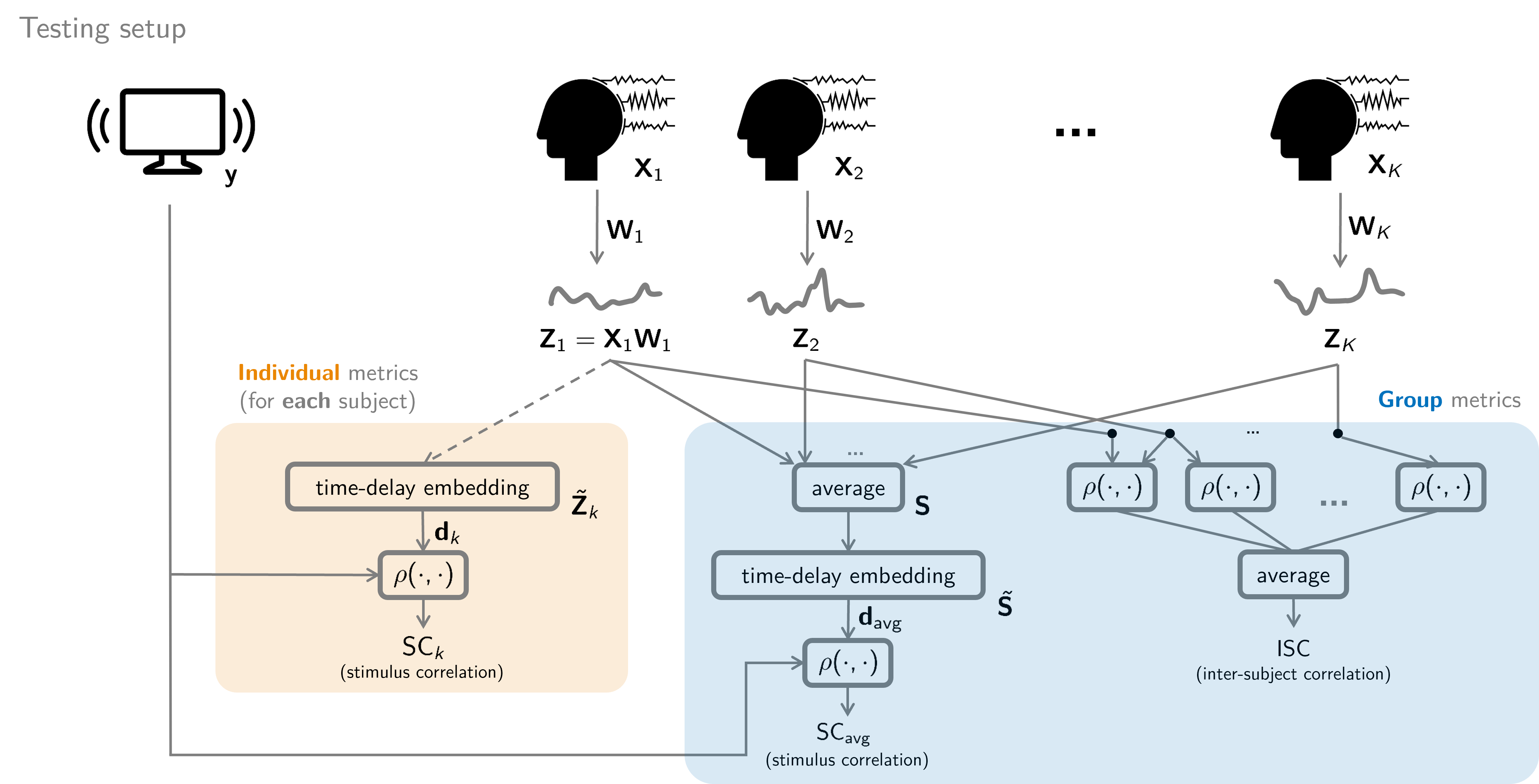}
		\caption{To compare the different methods, we use three evaluation metrics: one individual metric that quantifies the synchrony between each individual GCCA-decoded signal and the stimulus (i.e., the individual stimulus correlation $\text{SC}_k$ per subject $k$), and two group metrics that quantify the synchrony between the group summary GCCA-decoded signal and the stimulus (i.e., the stimulus correlation $\text{SC\textsubscript{avg}}$ on the average subspace signal) and the average synchrony between GCCA-decoded EEG signals (i.e., the inter-subject correlation (ISC)).}
		\label{fig:metrics}
	\end{figure*}

	\noindent 
	We consider two different evaluation metrics to compare the different methods: the inter-subject correlation (ISC) and stimulus correlation (SC). All evaluation metrics are schematically explained in Figure~\ref{fig:metrics}.
	
	\paragraph{Inter-subject correlation (ISC)}
	\label{sec:isc}
	\noindent
	GCCA is often used to quantify the group attention to a specific natural stimulus by using the ISC as a proxy for attentional engagement~\cite{dmochowski2012correlated,dmochowski2014audience,ki2016attention,poulsen2017eeg,yao2023identifying}. This ISC is defined as the average pairwise correlation coefficient between the GCCA-decoded EEG recordings of the different subjects, where the Pearson correlation coefficient $\rho\!\left(\vec{x},\vec{y}\right)$ between two zero-mean one-dimensional time signals $\vec{x} \in \R^T$ and $\vec{y} \in \R^T$ is defined as:
	
	\[
	\rho\!\left(\vec{x},\vec{y}\right) = \frac{\transpose{\vec{x}}\vec{y}}{\norm{\vec{x}}_2\norm{\vec{y}}_2}.
	\]
	Given the zero-mean one-dimensional projected EEG signals $\vec{z}_k^{(q)} = \mat{X}_k\vec{w}_k^{(q)} \in \R^{T}$ where $\vec{w}_k^{(q)}$ denotes the $q$\textsuperscript{th} column of the \mbox{(SI-)GCCA} decoder $\mat{W}_k$ (or of $\mat{W}$ in the case of corrCA), the ISC for component $q$ is thus defined as:
	\begin{equation}
		\label{eq:isc}
		\text{ISC}^{(q)} = \frac{2}{K(K-1)}\sumlim{k=1}{K-1}\sumlim{l=k+1}{K}\rho\!\left(\vec{z}_k^{(q)},\vec{z}_l^{(q)} \right).
	\end{equation}
	
	This ISC can be evaluated per window of a certain length on the test set to compare the various methods. Note that even for the stimulus-informed algorithms, at test time, we do not take the stimulus into account but compare them with the stimulus-unaware versions on exactly the same basis by only taking the projected EEG signals $\vec{z}_k^{(q)}$ into account.
	
	\paragraph{Stimulus correlation (SC)}
	\label{sec:sc}
	\noindent
	Similarly to \cite{deCheveigne2019multiway,diLiberto2021accurate}, we can also use \mbox{(SI-)GCCA}/corrCA as a preprocessing tool to a priori enhance the SNR of the stimulus-following responses by leveraging the group information, to then perform a traditional backward stimulus correlation analysis on the projected EEG signals $\mat{Z}_k = \mat{X}_k\mat{W}_k \in \R^{T \times Q}$. We can assume the stimulus feature $\vec{y} \in \R^T$ to be one-dimensional here (see Section~\ref{sec:stimulus-features}). The SC can then be found by first training a least-squares decoder $\vec{d}_k \in \R^{QL_d}$ from the time-lagged (using $L_d$ additional time lags) projected EEG signals $\tilde{\mat{Z}}_k \in \R^{T \times QL_d}$ to the stimulus $\vec{y}$. As explained in~\cite{geirnaert2021unsupervised}, this decoder can be found by solving the normal equations:
	\begin{equation}
		\label{eq:stim-dec}
		\vec{d}_k = \inverse{\left(\transpose{\tilde{\mat{Z}}_k}\tilde{\mat{Z}}_k\right)}\transpose{\tilde{\mat{Z}}_k}\vec{y}.
	\end{equation}
	The SC for subject $k$ can then be evaluated per window of a certain length on the test set by first applying the \mbox{(SI-)GCCA} decoders as a dimensionality reduction/preprocessing step and then applying the stimulus decoder:
	\begin{equation}
		\label{eq:sc}
		\text{SC}_k = \rho\!\left(\vec{y},\tilde{\mat{Z}}_k\vec{d}_k\right).
	\end{equation}
	
	While \eqref{eq:sc} contains the per-subject SC$_k$, we can also evaluate the overall SC across all subjects based on the subspace signal $\mat{S} = \frac{1}{K}\sumlim{k=1}{K}\mat{X}_k\mat{W}_k$ defined as the average projected EEG signal (similar to \eqref{eq:s}), which then acts as the summary signal. Using a decoder
	\begin{equation}
		\label{eq:stim-dec-avg}
		\vec{d}_{\text{avg}} = \inverse{\left(\transpose{\tilde{\mat{S}}}\tilde{\mat{S}}\right)}\transpose{\tilde{\mat{S}}}\vec{y}
	\end{equation}
	trained on the time-lagged average subspace signal $\tilde{\mat{S}} \in \R^{T \times QL_d}$ of the training set, the resulting stimulus correlation SC\textsubscript{avg} then represents a proxy for the group attention:
	\begin{equation}
		\label{eq:sc_avg}
		\text{SC\textsubscript{avg}} = \rho\!\left(\vec{y},\tilde{\mat{S}}\vec{d}_{\text{avg}}\right).
	\end{equation}
	
	These stimulus decoders \eqref{eq:stim-dec} and \eqref{eq:stim-dec-avg} are trained on the same data (i.e., the training set) on which the \mbox{\mbox{(SI-)GCCA}}/corrCA decoders are trained. That means the \mbox{(SI-)GCCA}/corrCA decoders are first applied to the training set, after which the decoders in \eqref{eq:stim-dec} and \eqref{eq:stim-dec-avg} can be trained. 
	
	\subsubsection{Significance level computation}
	The significance levels of the ISC and SC are computed using a random permutation test, where all correlation between the data is removed by randomly permuting all trials of the different subjects w.r.t. each other or randomly permuting the (projected) EEG and stimulus trials w.r.t. each other. Firstly, 50 Monte Carlo runs are performed in which the 40 training trials are randomly selected and the GCCA and stimulus decoders are trained. At test time, for each Monte Carlo run, 20 random permutations of the test trials are conducted, leading to a total of 1000 resamplings to determine the null distribution of the ISC and SC (i.e., when all correlated patterns are removed). From these null distributions, the $5\%$-significance level can then be computed and is the same for every algorithm. This significance level of the correlations is mainly determined by the window length ($\SI{60}{\second}$) and the number of subjects/group size (which is varied from 2 to the maximal number of subjects in the dataset) used to compare the correlation. 
	
	\subsection{Decoder setup}
	\label{sec:decoder-setup}
	\noindent
	
	\subsubsection{Filter design}
	In both the speech and video case, the neural decoders are modeled using a spatiotemporal filter that linearly combines the different EEG channels on different time lags (see Section~\ref{sec:maxvar-gcca}). The number of time lags $L$ is chosen equal to 5, corresponding to an integration window of $[-250,250]\SI{}{\milli\second}$ for speech, whereas in the video case, this corresponds to $[-66.7,66.7]\SI{}{\milli\second}$ (in accordance with~\cite{yao2023identifying}).
	
	The additional stimulus decoder to compute the SC in Section~\ref{sec:sc} is similarly modeled as a spatio-temporal decoder, but where all time lags are now chosen post-stimulus, i.e., from 0 to $L_d-1$ \emph{after} the current stimulus sample. In both cases, the integration window is chosen equal to $[0,250]\SI{}{\milli\second}$ (post-stimulus), corresponding to $L_d = 3$ in the speech case and $L_d = 9$ in the video case, consistent with, e.g., \cite{osullivan2014attentional,geirnaert2021eegBased}.
	
	In the SI-GCCA/-corrCA estimation, the stimulus representation is augmented with additional time-lagged copies to also allow for temporal filtering (via $\mat{V}$) at the stimulus side (including compensation for the intrinsic delay between the stimulus and EEG response), resulting in a Hankel matrix $\mat{Y}$. In the speech case, the integration window is chosen equal to $[-1.25,0]\SI{}{\second}$, i.e., preceding the current sample (and response), resulting in a Hankel matrix with $P = 11$ columns. In the video case, the integration window is chosen consistent with~\cite{yao2023identifying} equal to $[-500,0]\SI{}{\milli\second}$, resulting in a Hankel matrix with $P = 16$ columns.
	
	In the speech case, maximally $Q = 32$ components are extracted, whereas this is $Q = 10$ in the video case.
	
	\subsubsection{Hyperparameter selection}
	\label{sec:hyperparameter-selection}
	The hyperparameter $\mu$ in the GCCA \eqref{eq:maxvar-gcca-opt} and corrCA \eqref{eq:maxvar-corrCA-opt} estimation problems, determining the weight on the diagonal loading regularization of the neural decoders, and the hyperparameter $\gamma$ in the SI-GCCA \eqref{eq:sigcca-opt} and SI-corrCA \eqref{eq:sicorrca-opt} estimation problems, determining the weight on the stimulus part, are selected based on the average ISC on a validation set, independent from the training and test set. The optimal hyperparameter is selected based on the maximal average ISC for the first component across the $\SI{1}{\minute}$-trials in the validation set. For $\mu$, a sweep in the range of $\{0,10^{-5},10^{-4.5},\dots,10^5\}$ is performed. For $\gamma$ a sweep in the range of $\{0,10^{-2},10^{-1.5},\dots,10^8\}$ is performed.
	
	To not further complicate the hyperparameter search in the speech case, in the SI-GCCA \eqref{eq:sigcca-opt} and SI-corrCA \eqref{eq:sicorrca-opt} estimation problems, the hyperparameter $\mu$ for diagonal loading is not validated, but automatically (heuristically) determined using the method suggested by Ledoit and Wolf~\cite{ledoit2004well}. In the video case, the optimal validated $\mu$ from the GCCA problem is used in the SI-GCCA problem, as the Ledoit-Wolf procedure did not lead to satisfying results in this case. 
	
	\section{Results and discussion}
	\label{sec:results-discussion}
	\noindent
	We compare GCCA/corrCA with and without diagonal loading from Section~\ref{sec:maxvar-gcca/corrca} with the newly proposed stimulus-informed counterparts from Section~\ref{sec:si-gcca/corrca} both on the speech and video dataset according to the experiment details from Section~\ref{sec:experiments}. First, we consider only the speech dataset, varying three variables: the amount of available training data (Section~\ref{sec:atd}), the group size (Section~\ref{sec:gs}), and the number of EEG channels (Section~\ref{sec:nbc}). In Section~\ref{sec:interaction}, we investigate how the interaction of these three variables can influence the comparison. While the former experiments are driven from the perspective of using the stimulus-informed version to help the estimation of the neural decoders, in Section~\ref{sec:steering}, we take the alternative perspective of using SI-GCCA to steer the estimation of components towards the chosen stimulus representation.
	
	In Section~\ref{sec:analysis-video}, we then investigate and explain the performance on the video dataset.
	
	\subsection{Amount of training data}
	\label{sec:atd}
\begin{figure*}
		\centering
		\begin{subfigure}{0.35\textwidth}
			\renewcommand{\thesubfigure}{a-i}
			\includegraphics[width=\textwidth]{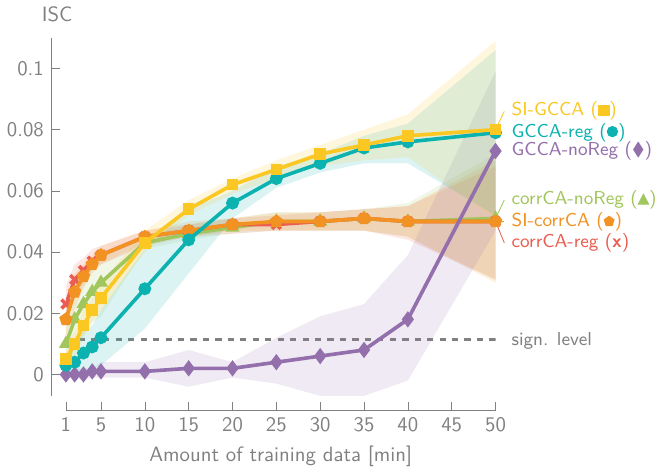}
			\caption{}
			\label{fig:isc-atd}
		\end{subfigure}%
		\begin{subfigure}{0.3\textwidth}
			\renewcommand{\thesubfigure}{a-ii}
			\includegraphics[width=\textwidth]{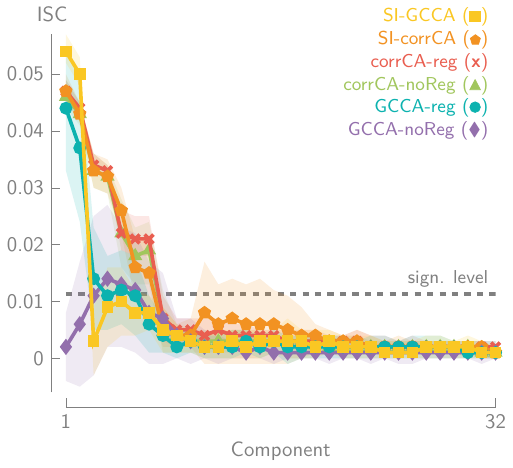}
			\caption{}		
			\label{fig:isc-comp-atd15}		
		\end{subfigure}%
		\begin{subfigure}{0.35\textwidth}
			\renewcommand{\thesubfigure}{a-iii}
			\includegraphics[width=\textwidth]{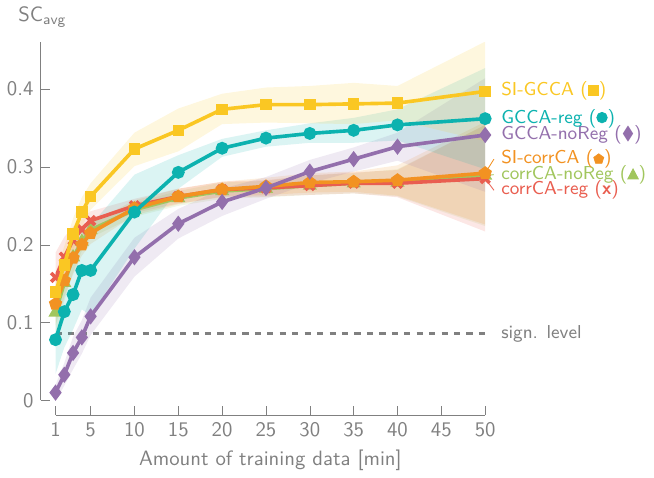}
			\caption{}		
			\label{fig:sc-atd-avgSub-cumul}		
		\end{subfigure}
		
		\begin{subfigure}{0.35\textwidth}
			\renewcommand{\thesubfigure}{b-i}
			\includegraphics[width=\textwidth]{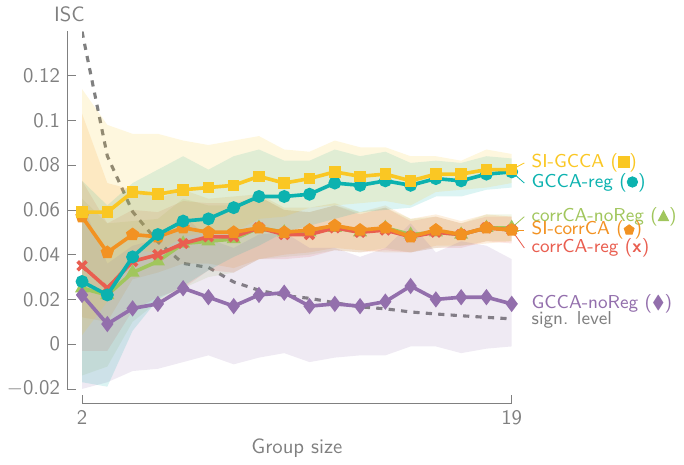}
			\caption{}
			\label{fig:isc-gs}
		\end{subfigure}%
		\begin{subfigure}{0.3\textwidth}
			\renewcommand{\thesubfigure}{b-ii}
			\includegraphics[width=\textwidth]{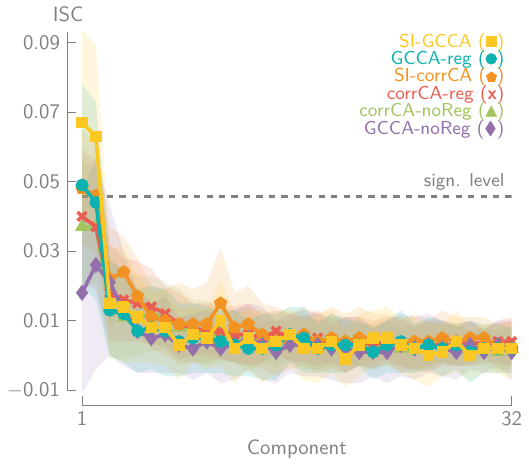}
			\caption{}		
			\label{fig:isc-comp-gs5}		
		\end{subfigure}%
		\begin{subfigure}{0.35\textwidth}
			\renewcommand{\thesubfigure}{b-iii}
			\includegraphics[width=\textwidth]{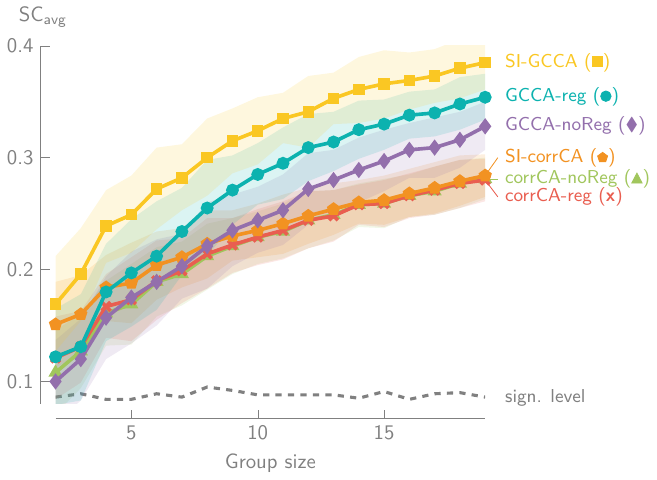}
			\caption{}		
			\label{fig:sc-gs-avgSub-cumul}		
		\end{subfigure}
		
		\begin{subfigure}{0.35\textwidth}
			\renewcommand{\thesubfigure}{c-i}
			\includegraphics[width=\textwidth]{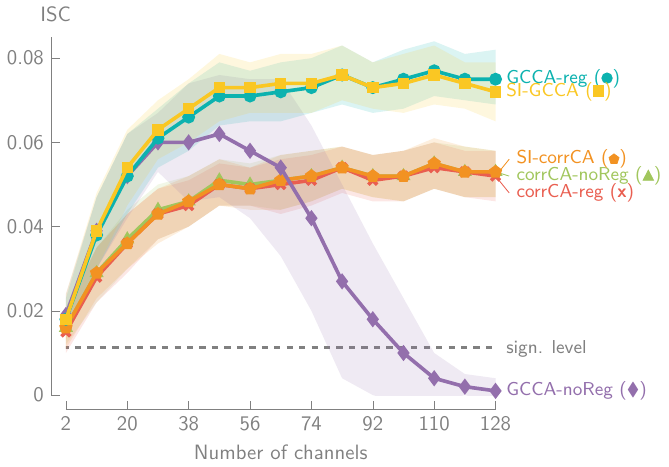}
			\caption{}
			\label{fig:isc-nbc}
		\end{subfigure}%
		\begin{subfigure}{0.3\textwidth}
			\renewcommand{\thesubfigure}{c-ii}
			\includegraphics[width=\textwidth]{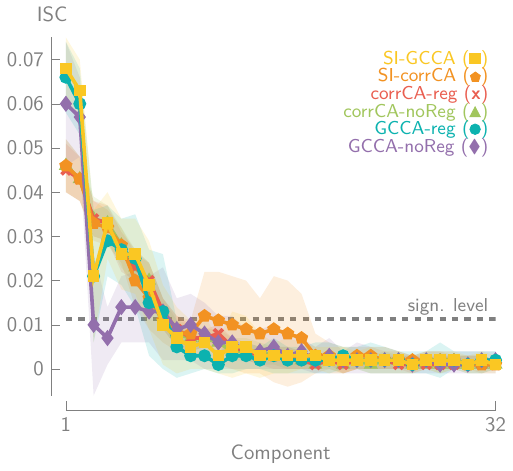}
			\caption{}		
			\label{fig:isc-comp-nbc38}		
		\end{subfigure}%
		\begin{subfigure}{0.35\textwidth}
			\renewcommand{\thesubfigure}{c-iii}
			\includegraphics[width=\textwidth]{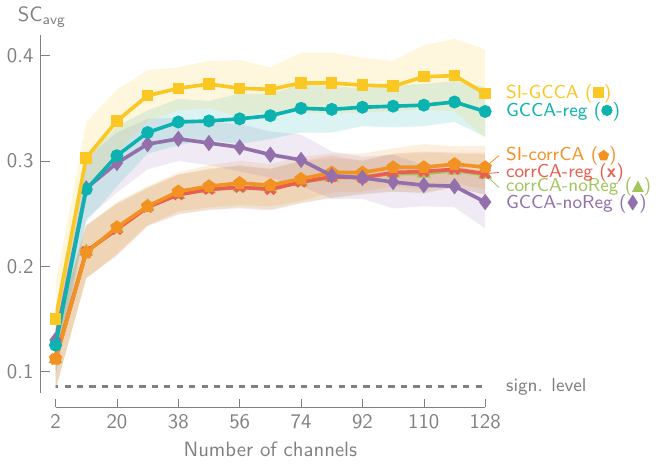}
			\caption{}		
			\label{fig:sc-nbc-avgSub-cumul}		
		\end{subfigure}
		
		\caption{(-i) The ISC as a function of the amount of training data (a-i), group size (b-i), and number of channels (c-i) for the first component on the speech data, when using 64 channels and all 19 subjects (mean $\pm$ standard deviation across runs). (a-i) For smaller amounts of training data, the SI-GCCA algorithm outperforms the GCCA algorithms, whereas the corrCA versions perform best for extremely low amounts of training data. (b-i) The stimulus-informed versions outperform the uninformed traditional versions for smaller group sizes. (c-i) GCCA clearly overfits without any regularization. (-ii) The ISC across components, for $\SI{15}{\minute}$ of training data (a-ii), group size 5 (b-ii), and 38 (random) channels (c-ii), showing mainly an effect of the inclusion of the stimulus on the most significant components. (-iii) The SC\textsubscript{avg} of the average subspace with $Q = 32$ components as a function of the amount of training data (a-iii), group size (b-iii), and number of channels (c-iii), showing how the stimulus information leads to a better preprocessor (a-iii and b-iii) and again how GCCA without any regularization overfits when using many EEG channels (c-iii).}
		\label{fig:overview-fig}
	\end{figure*}

	\noindent
	In this experiment, we vary the amount of available training data on the speech dataset from $\SI{1}{\minute}$ to $\SI{50}{\minute}$ while keeping the other variables fixed, as explained in Section~\ref{sec:performance-evaluation}. Per amount of training data, 50 Monte Carlo runs of randomly picking the training trials are performed. Investigating different (smaller) amounts of training data is especially relevant, for example, in a time-adaptive context, when the decoder has to be updated regularly to cope with the non-stationarities in the data~\cite{geirnaert2022timeAdaptive}. Figure~\ref{fig:isc-atd} shows the ISC as a function of the amount of training data only for the first component, whereas Figure~\ref{fig:isc-comp-atd15} shows the ISC per component when using $\SI{15}{\minute}$ of training data. Figure~\ref{fig:sc-atd-avgSub-cumul} shows the SC\textsubscript{avg} of the average decoded subspace $\mat{S}$ as a function of the amount of training data when using all $Q = 32$ components. The no regularization (`noReg') case refers to $\mu = 0$ in \eqref{eq:maxvar-gcca-opt} (for GCCA) and \eqref{eq:maxvar-corrCA-opt} (for corrCA), whereas the regularized case (`reg') refers to the case with $\mu$ selected based on the validation set performance. 
	
	A first important observation from Figure~\ref{fig:isc-atd}, and from Figure~\ref{fig:overview-fig} in general is that significant ISCs can be obtained using data-driven group decoding algorithms. This shows that synchronized stimulus-following responses across subjects exist and can be decoded. The latter has already been established in other studies on neural responses to natural stimuli~\cite{dmochowski2012correlated,ki2016attention,poulsen2017eeg}, and the ISC values in Figure~\ref{fig:overview-fig} are in line with those observed in other studies. In general, ISC and SC values are typically low because of the very low SNR of the stimulus-following neural responses w.r.t. the background EEG activity.
	
	From Figure~\ref{fig:isc-atd}, we learn that SI-GCCA outperforms GCCA, especially when a smaller amount of training data is available. Firstly, it can be seen that traditional GCCA without any regularization or side-information heavily suffers from overfitting in this case: only when more than $\SI{35}{\minute}$ of training data are available, a significant ISC is found. With diagonal loading, this overfitting effect can be counteracted by using the prior information that norms of the neural decoders should be limited, leading to a significant ISC already with $\SI{5}{\minute}$ of training data. However, a smarter way of introducing side information seems to be using the stimulus as proposed in the SI-GCCA algorithm: especially for smaller amounts of training data, this outperforms GCCA, leading to significant ISCs for $\SI{3}{\minute}$ of training data. Note that from a numerical perspective, the $\ell_2$-norm regularization and stimulus information are different: in the former, no additional parameters need to be estimated when using this prior information, whereas in SI-GCCA, additional parameters are introduced in the problem. When increasing the amount of training data, the difference between SI-GCCA and GCCA versions becomes smaller as the additional side-information introduced by the stimulus is outweighed by the large amount of training data, effectively compensating for the high dimensionality of the problem.
	
	Another effective way of coping with the smaller amount of training data is by drastically decreasing the dimensionality of the estimation problem as in the \mbox{(SI-)corrCA} problems. However, there seems to be no additional benefit from the SI-corrCA method w.r.t. corrCA with diagonal loading for extremely low amounts of training data. The corrCA regularization technique proves to be very effective for very low amounts of training data: below $\SI{10}{\minute}$, it outperforms SI-GCCA. However, the flip side is that the performance quickly saturates when the amount of training data increases. \mbox{(SI-)corrCA} clearly suffers from its inability to model subject differences w.r.t. \mbox{(SI-)GCCA}, leading to vastly lower ISCs.
	
	Figure~\ref{fig:isc-comp-atd15} additionally shows that SI-GCCA mainly boosts the ISC for the most significant components, whereas in \mbox{(SI-)corrCA}, more significant components can be found - the ISC is more spread out. In principle, the components cannot be compared between methods one-to-one, as they might represent different activities and only jointly form the basis for the subspace. Therefore, Figure~\ref{fig:sc-atd-avgSub-cumul} makes it easier to compare across different components, as here the average subspaces across methods are compared in terms of their SC\textsubscript{avg}. As expected because of the design of the SI-GCCA method, it is more effective as a preprocessing tool for stimulus decoding, yielding higher stimulus correlations. 
	
	\subsection{Group size}
	\label{sec:gs}
	\noindent
	In this experiment, we vary the group size from 2 to 19 on the speech dataset while keeping the other variables fixed. Similarly as before, 50 Monte Carlo runs of randomly picking the training trials \emph{and} the group of subjects for a specific group size are performed. This experiment emulates situations where the group size is limited.
	
	%		\begin{figure*}
		%			\centering
		%			\begin{subfigure}{0.35\textwidth}
			%				\includegraphics[width=\textwidth]{plots/isc-gs}
			%				\caption{}
			%				\label{fig:isc-gs}
			%			\end{subfigure}%
		%			\begin{subfigure}{0.3\textwidth}
			%				\includegraphics[width=\textwidth]{plots/isc-comp-gs5}
			%				\caption{}		
			%				\label{fig:isc-comp-gs5}		
			%			\end{subfigure}%
		%			\begin{subfigure}{0.35\textwidth}
			%				\includegraphics[width=\textwidth]{plots/sc-gs-avgSub-cumul}
			%				\caption{}		
			%				\label{fig:sc-gs-avgSub-cumul}		
			%			\end{subfigure}
		%			\caption{(a) The ISC as a function of the group size for the first component on the speech data (mean $\pm$ standard deviation across runs). The stimulus-informed versions outperform the uninformed traditional versions for smaller group sizes. (b) The ISC for group size 5 across components, showing mainly an effect of the inclusion of the stimulus on the most significant components. (c) The SC\textsubscript{avg} of the average subspace with $Q = 32$ components as a function of group size, showing how the stimulus information leads to a better preprocessor.}
		%			\label{fig:gs}
		%		\end{figure*}
	
	Figure~\ref{fig:isc-gs} shows an even clearer effect of the stimulus information w.r.t. the uninformed counterparts when the group size is limited, not only for SI-GCCA but now also for SI-corrCA. The stimulus side-information effectively compensates when less information is available due to a smaller group size, leading to a significant\footnote{Note that the significance level decreases with an increasing group size, as when more subjects are available, more pairwise correlations are averaged in the ISC \eqref{eq:isc}, suppressing more potential spurious correlations.} ISC already for 3 subjects in the case of SI-GCCA and only a minor decrease in ISC. Moreover, the corrCA variants are now outperformed each time by a GCCA counterpart, as sufficient ($\SI{40}{\minute}$) training data is available (see Figure~\ref{fig:isc-atd}). 
	
	In Figure~\ref{fig:isc-comp-gs5}, it can again be seen that the stimulus information mainly boosts the most significant components, while SI-GCCA/SI-corrCA are also more effective preprocessing tools w.r.t. the stimulus-uninformed traditional versions, as seen in Figure~\ref{fig:sc-gs-avgSub-cumul}.
	
	\subsection{Number of channels}
	\label{sec:nbc}
	\noindent
	In this experiment, we vary the number of channels from 2 to 128 in steps of 9 on the speech dataset. Similarly as before, 50 Monte Carlo runs of randomly picking the training trials \emph{and} the EEG channels (randomly chosen) for a specific number of channels are performed. As such, we vary the number of parameters to be estimated \emph{and} the available information.
	
	%		\begin{figure*}
		%			\centering
		%			\begin{subfigure}{0.35\textwidth}
			%				\includegraphics[width=\textwidth]{plots/isc-nbc}
			%				\caption{}
			%				\label{fig:isc-nbc}
			%			\end{subfigure}%
		%			\begin{subfigure}{0.3\textwidth}
			%				\includegraphics[width=\textwidth]{plots/isc-comp-nbc38}
			%				\caption{}		
			%				\label{fig:isc-comp-nbc38}		
			%			\end{subfigure}%
		%			\begin{subfigure}{0.35\textwidth}
			%				\includegraphics[width=\textwidth]{plots/sc-nbc-avgSub-cumul}
			%				\caption{}		
			%				\label{fig:sc-nbc-avgSub-cumul}		
			%			\end{subfigure}
		%			\caption{(a) The ISC as a function of the number of channels for the first component on the speech data (mean $\pm$ standard deviation across runs), clearly showing overfitting for GCCA without any regularization. There is no difference between SI-GCCA/corrCA and GCCA/corrCA with diagonal loading. (b) The ISC for 38 (random) channels. (c) The SC\textsubscript{avg} of the average subspace with $Q = 32$ ($Q = 10$ for only 2 channels) components as a function of number channels, again showing how GCCA without any regularization overfits when using many EEG channels.}
		%			\label{fig:nbc}
		%		\end{figure*}
	
	Figure~\ref{fig:isc-nbc} and \ref{fig:sc-nbc-avgSub-cumul} very clearly show the effect of overfitting on the GCCA method when the number of channels increases, when not using any regularization method or side-information. However, there seems to be no difference between the stimulus-informed and -uninformed (but regularized with diagonal loading) versions in Figure~\ref{fig:isc-nbc} and \ref{fig:isc-comp-nbc38} when varying the number of channels. While SI-GCCA results in a slight improvement w.r.t. GCCA for the SC (Figure~\ref{fig:sc-nbc-avgSub-cumul}), the stimulus seems to not help in this situation in terms of the ISC, especially when the other variables (amount of training data and group size) are equal to the (relatively large) default values. However, in Section~\ref{sec:interaction}, we investigate what happens when these three variables interact, and we will show that the number of channels can have an influence when less ideal variable values are used.
	
	\subsection{Interaction of variables}
	\label{sec:interaction}
	\noindent
	In the previous experiments, only one variable is varied each time, while the other variables are taken constant at their default values as explained in Section~\ref{sec:performance-evaluation}. As such, the previous results are, in a certain sense, still quite conservative, as two of the three variables are each time taken to be quite ideal in the sense that they already lead to data-rich settings, even when the third variable is set to a low value. In this section, we explore how these variables interact and influence the comparison by taking less ideal values for all three variables, which can easily occur in a practical application. More specifically, as a representative example, we choose the amount of training data equal to $\SI{15}{\minute}$, the group size equal to 6, and vary the number of channels from 2 to 65 in steps of 9 on the speech dataset.
	
\begin{figure}
		\centering
		\includegraphics[width=0.38\textwidth]{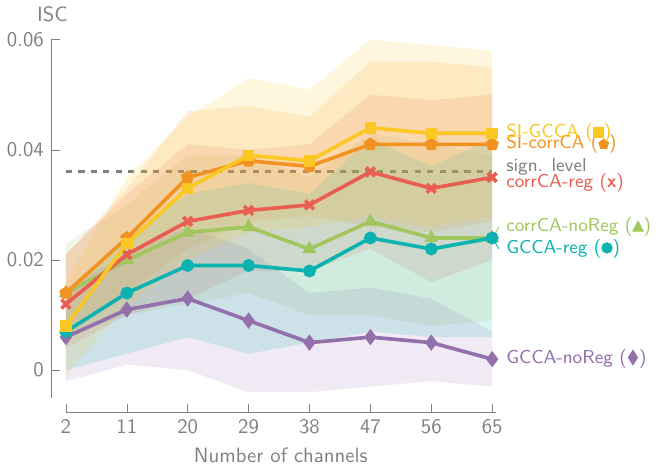}
		\caption{Both SI-GCCA and SI-corrCA clearly outperform their uninformed counterparts across various numbers of channels when only $\SI{15}{\minute}$ of training data and 6 subjects are available (first component).}
		\label{fig:isc-interaction}
	\end{figure}

	Figure~\ref{fig:isc-interaction} shows how this interaction between variables favors even more the stimulus-informed versions w.r.t. the stimulus-uninformed GCCA variants across multiple numbers of channels. Only when using the stimulus, a significant ISC can be obtained when using more than 30 channels. Moreover, similarly to small group sizes in Figure~\ref{fig:isc-gs}, SI-corrCA substantially outperforms corrCA with diagonal loading. This particular instance showcases how in non-ideal, practical use cases, where the amount of training data \emph{and} the group size is limited, our newly proposed SI-GCCA/corrCA algorithms can lead to a substantial benefit.
	
	\subsection{Steering the GCCA estimation}
	\label{sec:steering}
	\noindent
	In Sections \ref{sec:atd} to \ref{sec:interaction}, we have shown how SI-GCCA is superior when the available information to estimate the correlated components is limited, e.g., because the amount of training data or group size is limited. Interpreted broadly (not numerically), SI-GCCA can be seen as a task-informed regularization technique that allows to introduce additional available information when estimating correlated components from stimulus-synchronized EEG activity. In this section, we want to put forward an alternative interpretation of SI-GCCA, i.e., using the stimulus to steer the estimation of the correlated components towards the stimulus. This alternative perspective connects with employing \mbox{(SI-)GCCA}, e.g., as a preprocessor for stimulus decoding~\cite{diLiberto2021accurate} or to use the ISC as a proxy for attentional engagement to the content of one particular stimulus, for example, the teacher's voice in a classroom~\cite{poulsen2017eeg}.
	
	From Figure~\ref{fig:isc-nbc} and \ref{fig:sc-nbc-avgSub-cumul}, it can already be seen that for a similar ISC, SI-GCCA can still lead to higher SC w.r.t. GCCA. This already indicates that while the extracted components per subject are almost equally well correlated with one another, the ones extracted by SI-GCCA are still more related to the stimulus (feature) than for GCCA, in other words, SI-GCCA steers these components more towards the stimulus. Whereas Figure~\ref{fig:sc-nbc-avgSub-cumul} shows only the SC for the average subspace, we further investigate whether this steering behavior is also present on an individual level, per component and subject. Therefore, we more closely compare SI-GCCA and GCCA with diagonal loading for $\SI{40}{\minute}$ of training data, using all 19 subjects and the default 64 EEG channels. 
	
	\begin{figure}
		\centering
		%	\begin{subfigure}{0.33\textwidth}
			%		\includegraphics[width=\textwidth]{plots/isc-comp-steering}
			%		\caption{}
			%		\label{fig:steering-isc}
			%	\end{subfigure}%
		%	\begin{subfigure}{0.33\textwidth}
			%		\includegraphics[width=\textwidth]{plots/sc-comp-avgSub}
			%		\caption{}		
			%		\label{fig:steering-sc-comp}		
			%	\end{subfigure}%
		\begin{subfigure}{0.5\textwidth}
			\centering
			\includegraphics[width=0.92\textwidth]{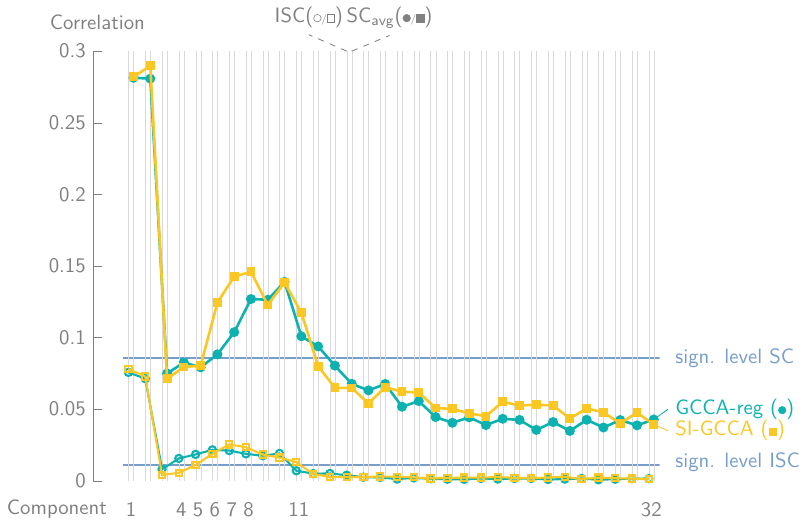}
			\caption{}
			\label{fig:steering-isc-sc}
		\end{subfigure}
		
		\begin{subfigure}{0.5\textwidth}
			\centering
			\includegraphics[width=0.8\textwidth]{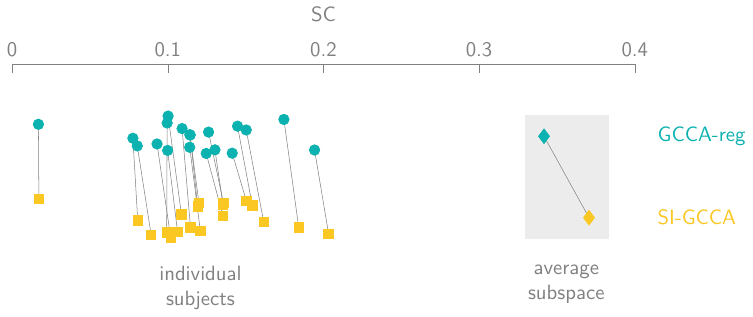}
			\caption{}		
			\label{fig:steering-sc-subj}		
		\end{subfigure}%
		\caption{(a) The ISC between SI-GCCA and GCCA with diagonal loading is very similar across components when using $\SI{40}{\minute}$ of training data, all 19 subjects, and the default 64 channels. Mostly for components 6 to 8, SI-GCCA leads to a higher SC\textsubscript{avg} than GCCA. (b) For all individual subjects, the SC when using the first 11 components extracted with \mbox{(SI-)GCCA} is higher for SI-GCCA than GCCA (Wilcoxon signed-rank test: $n = 19$, $p$-value $<0.001$). This effect is even more amplified when working with the 11-dimensional average subspace.}
		\label{fig:steering}
	\end{figure}
	
	As could be seen already in Figure~\ref{fig:isc-atd} for the default settings and is now confirmed in Figure~\ref{fig:steering-isc-sc}, especially for the most significant components, there is hardly any difference in terms of ISC. When summing the ISCs across all components, GCCA and SI-GCCA lead to the same cumulative ISC of $0.335$, and no significant difference can be found across components (Wilcoxon signed rank test, $n = 32$, $p$-value$=0.24$). However, when we have a look at the SC per individual component in Figure~\ref{fig:steering-isc-sc}, SI-GCCA almost always leads to a higher or comparable SC than GCCA, even when the ISC is lower (e.g., for component 4, 5, 6). This is specifically noticeable for components 6 to 8. Moreover, in Figure~\ref{fig:steering-sc-subj}, we show the SC when using the 11-dimensional subspace (all components beyond the 11\textsuperscript{th} are not significant in Figure~\ref{fig:steering-isc-sc}) from \mbox{(SI-)GCCA}, per individual subject and also for the average subspace. For all subjects, the SC is higher for the SI-GCCA method compared to the GCCA method, showing a significant improvement across subjects (Wilcoxon signed-rank test: $n = 19$, $p$-value $<0.001$). Furthermore, this effect is amplified when using the average subspace. This shows how SI-GCCA can also be used to steer the correlated components to be more correlated with the stimulus and shows its benefit as a preprocessor to boost the SNR before stimulus decoding, both on an individual and group level.
	
	\subsection{Analysis of video data with object-based optical flow and the effect of the specific stimulus representation}
	\label{sec:analysis-video}
	\noindent
	In this section, we evaluate and compare SI-GCCA with GCCA on the video dataset. 
	
\begin{figure}
		\centering
		\begin{subfigure}{0.33\textwidth}
			\centering
			\includegraphics[width=\textwidth]{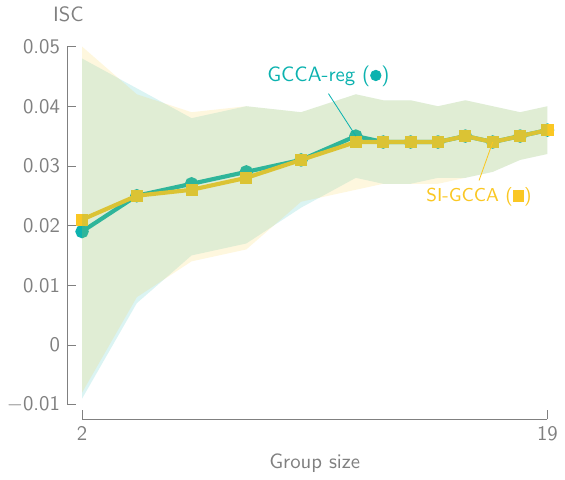}
			\caption{}
			\label{fig:video-isc-gs}
		\end{subfigure}
		
		\begin{subfigure}{0.44\textwidth}
			\includegraphics[width=\textwidth]{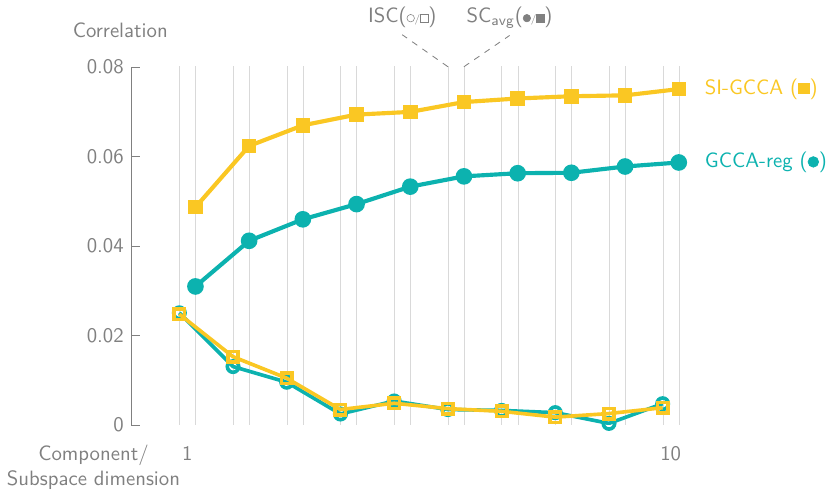}
			\caption{}		
			\label{fig:video-isc-sc-comp}		
		\end{subfigure}%
		%	\begin{subfigure}{0.33\textwidth}
			%		\includegraphics[width=\textwidth]{plots/scavg-comp-video}
			%		\caption{}		
			%		\label{fig:video-scavg}		
			%	\end{subfigure}%
		\caption{(a) The ISC as a function of group size with $\SI{40}{\minute}$ of training data and 64 EEG channels for the video dataset. There is no benefit from the stimulus-informed version. (b) There is hardly any difference when comparing the ISC between SI-GCCA and GCCA across components for group size 4. However, SI-GCCA clearly leads to a higher SC\textsubscript{avg} using increasing subspace dimension, showing its capability to steer the estimation of the correlated components towards the stimulus representation.}
		\label{fig:video}
	\end{figure}

	In Figure~\ref{fig:video-isc-gs}, the ISC is shown as a function of group size when using $\SI{40}{\minute}$ of training data and all 64 EEG channels. From Figure~\ref{fig:isc-gs}, we would expect from the speech dataset that the stimulus information would especially help for group sizes below 10. However, Figure~\ref{fig:video-isc-gs} shows hardly any difference between SI-GCCA and GCCA, indicating that the stimulus does not improve the estimation of the correlated components (but also does not deteriorate it). An explanation can be found in Yao et al.~\cite{yao2023identifying}, where it is shown that the object-based optical flow only explains $6.9\%$ of the variance present in the correlated components across the EEG's of the different subjects. A large proportion of the variance in the stimulus-related neural responses is thus not yet explained, such that using this specific video feature in SI-GCCA has only a minor effect on the estimated correlated components in terms of maximizing the ISC. To strengthen this conclusion, we have performed a similar experiment as in \cite{yao2023identifying} to quantify the variance explained in the GCCA components for the speech dataset when using the speech envelope as a feature. When regressing out the speech envelope per subject and re-estimating GCCA (more details in \cite{yao2023identifying}), we find that the speech envelope explains around $40.6\%$ of all correlated activity, explaining why SI-GCCA performs much better on the speech dataset. This observation entails an important limitation when using our proposed SI-GCCA method: its impact is bound to the specific chosen stimulus representation(s).
	
	However, even when the stimulus representation does not explain much of the correlated neural (stimulus-related) activity as is the case here for object-based optical flow in the video dataset, SI-GCCA can still be used to steer the correlated components towards that specific stimulus representation. We highlight this point by comparing the ISC on the video dataset for group size 4 with the SC\textsubscript{avg} across all components. While Figure~\ref{fig:video-isc-sc-comp} shows hardly any difference in ISC per individual correlated components, the subspaces of increasing dimension are clearly much more related to the object flow feature, already starting from a one-dimensional subspace. This improved SC does not reduce the correlation across subjects, as the ISC remains the same between both methods. This shows that SI-GCCA extracts equally correlated components as GCCA (across subjects), but the former extracts components that better capture the temporal dynamics in the optical flow related to the moving object in the video.
	
	\section{Conclusion}
	\label{sec:conclusion}
	\noindent
	In this paper, we proposed a new algorithm for the group analysis of stimulus-following neural responses within a group of people attending to the same natural stimulus. Our proposed framework allows to take the stimulus itself into account when estimating the correlated components across subjects with GCCA and its subject-generic variant corrCA. This stimulus-informed GCCA framework can still be solved as a GEVD, inheriting this attractive property from MAXVAR-GCCA.
	
	We compared our newly proposed SI-GCCA algorithms with the traditional stimulus-uninformed versions on a speech and video dataset, using the speech envelope and object-based optical flow as exemplary stimulus representations. We demonstrated the superiority of using the stimulus as side-information when the amount of training data or group size is limited, even more so when these different variables interact, also with the number of channels. This shows its practical relevance, for example, in situations where the training set size is limited, e.g., in a context of time-adaptive, online processing, or when the group size is limited, e.g., as determined by the application. Using the video dataset, we showed that a limitation of SI-GCCA, besides the requirement of having access to the stimulus, is its dependency on the specific stimulus representation.
	
	Besides using SI-GCCA to introduce the stimulus as valuable side-information to robustify the estimation of correlated components across a group of subjects when the available estimation data is a priori limited, it can also be used to steer the correlated components explicitly in the direction of the stimulus (features). To this end, we showed that a higher stimulus correlation can be obtained when using SI-GCCA versus the traditional uninformed GCCA, without any significant reduction in the ISC.
	
	To sum up, the proposed SI-GCCA algorithm can be employed for various purposes in the group decoding of stimulus-following neural responses, e.g., to avoid overfitting and compensate for limited available information, or to steer the design of the neural decoders towards a specific stimulus representation. As such, it can enable various applications, ranging from more fundamentally-oriented (e.g., preprocessing and dimensionality reduction) to application-specific (e.g., quantifying attention in the classroom).

	\bibliographystyle{ieeetran}
	\bibliography{papers-ref-si-gcca}
	
\end{document}